%% file: main.tex
\documentclass[final,3p,times]{elsarticle}
\usepackage{amssymb}
\usepackage{amsmath}
\usepackage{mathrsfs}
\usepackage{algorithm}
\usepackage{algpseudocode}
\usepackage{multirow}
\usepackage{url}
\usepackage{eso-pic}
\usepackage{graphicx}

\renewcommand{\vec}[1]{\mathbf{#1}}

\newcommand{\curl}{\mathrm{curl\,}}
\renewcommand{\div}{\mathrm{div\,}}

\journal{Journal of Computational Physics}

\begin{document}

\begin{frontmatter}

\title{Accounting for Hysteresis and Eddy Currents in Finite Element Simulations of Ferromagnetic Laminated Cores using a Recurrent Neural Network}
\author[1]{Florent Purnode\corref{cor1}}\ead{florent.purnode@uliege.be}
\author[1]{Louis Denis}
\author[1]{François Henrotte}
\author[1]{Gilles Louppe}
\author[1]{Christophe Geuzaine}\ead{cgeuzaine@uliege.be}
\cortext[cor1]{Corresponding author}
\affiliation[1]{
	organization={Department of Electrical Engineering and Computer Science, University of Liege},
	country={Belgium}
}

\begin{abstract}
	Incorporating hysteresis and eddy currents
	into finite element simulations
	of laminated-core electrical machines is computationally challenging.
	Resolving the fields inside the laminations
	at each integration point and at every nonlinear iteration
	leads to computational costs several orders of magnitude higher than anhysteretic simulations,
	making such approaches impractical for design applications.
	Conversely, simplified models accounting only for magnetic saturation
	are becoming increasingly inadequate
	as electrical machine topologies and operating conditions grow in complexity.
	In this context, machine learning surrogate modeling has emerged as a promising alternative,
	offering efficient and accurate approximations of complex electromagnetic behaviors.
	In this paper, a recurrent neural network is trained
	as a surrogate of a laminated-core material model for an isotropic laminated core,
	and is integrated into realistic two-dimensional magnetodynamic finite element simulations
	based on a magnetic vector potential formulation.
	The proposed approach achieves excellent agreement
	with the reference laminated-core model
	while limiting the computational cost to about twice
	that of an anhysteretic simulation.
	By training the recurrent neural network
	on a sufficiently diverse set of artificially generated magnetic field sequences
	designed to mimic those encountered in electrical machine simulations,
	the proposed approach can be readily applied across a wide range of finite element simulations.
	Furthermore, the trained surrogate model is provided as a standalone component
	that can be easily incorporated into existing computational frameworks.
	It is publicly available at \url{https://gitlab.onelab.info/getdp/lamnet}.
\end{abstract}

\AddToShipoutPicture*{
	\footnotesize\sffamily\raisebox{2.5cm}{
		\hspace{2cm}\fbox{
			\parbox{16.1cm}{
				This work has been submitted to a journal for possible publication. Copyright may be transferred without notice, after which this version may no longer be accessible.
			}
		}
	}
}


\end{frontmatter}

\section{Introduction}
\input{introduction}

\section{Lamination Model}
\label{sec:tgtModel}
\input{lam}

\section{Architecture and Training of the NN-Based Surrogate Model}
\label{sec:NN}
\input{NN}

\section{Macroscopic-Scale Simulation}
\label{sec:FEM}
\input{FEM}

\section{Conclusions and Perspectives}
\label{sec:conclusion}
\input{conclusion}

\appendix
\section{Parameters of the Energy-Based Model}
\label{sec:EBparams}
\begin{center}
	\begin{tabular}{l|ccccccccccc}
		\noalign{\hrule height 1.2pt}
		\noalign{\vskip 5pt}
		\multicolumn{2}{c|}{$h_a=18.18$~A/m}&\multicolumn{2}{c|}{$h_b=3905.7$~A/m}&\multicolumn{2}{c|}{$J_a=1.387$~T}&\multicolumn{2}{c|}{$J_b=0.559$~T}&\multicolumn{2}{c|}{$\rho = 690~\mathrm{n\Omega\,m}$} & \multicolumn{2}{c}{$d=350~\mu$m}\\
		\noalign{\vskip 5pt}
		\noalign{\hrule height 1.2pt}
		\noalign{\vskip 5pt}
		\multicolumn{12}{l}{$N=11$ elementary cells:}\\
		\noalign{\vskip 5pt}
		$\kappa^n~[\text{A/m}]$ & 0.0 & 7.3 & 18.8 & 32.1 & 45.5 & 55.8 & 66.9 & 80.6 & 99.1 & 143.0 & 213.5\\
		$w^n~[-]$ & 0.075 & 0.103 & 0.106 & 0.342 & 0.119 & 0.105 & 0.053 & 0.043 & 0.028 & 0.019 & 0.006\\
		\noalign{\vskip 5pt}
		\noalign{\hrule height 1.2pt}
	\end{tabular}
\end{center}

\section{Resampling algorithm}
\label{sec:resamplingAlg}
\input{resampling}

\section*{Acknowledgments}

Computational resources have been provided by the Consortium des Équipements de Calcul Intensif (CÉCI), funded by the Fonds de la Recherche Scientifique de Belgique (F.R.S.-FNRS) under Grant n\textsuperscript{o}2.5020.11 and by the Walloon Region.
The present research also benefited from computational resources
made available on Lucia, the Tier-1 supercomputer of the Walloon Region,
infrastructure funded by the Walloon Region under the grant agreement n\textsuperscript{o}1910247.

\section*{Funding sources}

This work was supported in part by the Walloon Region of Belgium
under grant 8139 M\&SSCoT and 9001 ELECTRON.
The work of Louis Denis was supported by the F.R.S-FNRS.

\section*{Declaration of generative AI and AI-assisted technologies in the manuscript preparation process}

During the preparation of this work,
the authors used ChatGPT in order to revise portions of the manuscript
and improve its language and readability.
After using this service,
the authors reviewed and edited the content as needed
and take full responsibility for the content of the published article.

\input{biblio.tex}
\end{document}

%% file: introduction.tex
Transformers and electric machines rely on laminated ferromagnetic cores
to guide and concentrate magnetic flux.
Accurate modeling of electromagnetic fields and magnetic losses
within these laminated regions
is crucial for the reliable prediction
of induced voltages and overall device efficiency
\cite{ceylan2023magnetodynamic, billah2020computationally}.
However, the electromagnetic behavior of laminated cores
is particularly challenging to model due to
the inhomogeneous distribution of eddy currents,
the nonlinear and history-dependent nature of magnetic hysteresis,
and the coupling between these two phenomena.

Modeling hysteresis alone is already challenging.
The Preisach~\cite{mayergoyz1988generalized}
and Jiles-Atherton~\cite{jiles1986theory} models
have long been considered standard hysteresis models,
and have been widely implemented in two-dimensional (2D)
finite element (FE) simulations
\cite{benabou2003comparison, li2014investigation, nierla2019comparison}.
However, both models were originally formulated for scalar hysteresis,
and their extensions to vector hysteresis lack a clear physical foundation
which limits their applicability to vector field problems.
In addition, they may suffer from limitations related to accuracy,
computational cost, or numerical convergence~\cite{atyia2024limitations, hussain2017efficient}.
More recently, the energy-based (EB) model~\cite{henrotte2006energy, Jacques}
has emerged as a promising alternative.
This hysteresis model is based on energy considerations
and is intrinsically vectorial~\cite{bergqvist1997magnetic}.
The forward EB model takes the magnetic field $\vec{H}$ as input
and returns the history-dependent magnetic flux density $\vec{B}$,
while its inverse formulation~\cite{egger2025forward, egger2025efficient},
computes $\vec{H}$ from a given $\vec{B}$.
The forward model can be evaluated either by solving a minimization
problem~\cite{franccois2013energy} or a nonlinear vector
function~\cite{Jacques}.
Besides these two formulations,
a simplified explicit variant,
akin to a vector play model~\cite{henrotte2006energy, Jacques},
also exists and provides a computationally efficient update
at the cost of a slightly reduced accuracy~\cite{roppert2025magneto}.
The inverse formulation, by contrast, requires solving a minimization
problem~\cite{Jacques, franccois2013energy, Kevin2016Magnetic},
whose robustness has been recently significantly improved \cite{egger2025forward, egger2025efficient}.

Eddy currents further add to the complexity of modeling ferromagnetic behavior.
They form closed current loops whose distribution is governed
by the spatial and temporal variations of the magnetic flux density.
In laminated cores, eddy currents are largely confined to the plane of each lamination,
but three-dimensional (3D) effects arise near edges and geometric discontinuities
and cannot be fully captured by the 2D formulations commonly employed in electrical machine design.
However, direct 3D modeling of laminated cores,
in which the fields are resolved within each individual lamination,
is several orders of magnitude more computationally expensive.
Although feasible on high-performance computing platforms,
such simulations remain impractical for engineering design and optimization.
The periodicity of the laminated structure can, however,
be exploited through homogenization techniques,
yielding 2D formulations that accurately capture the bulk eddy-current behavior,
although their accuracy deteriorates near edges and geometric discontinuities.
In multiscale methods
like the heterogeneous multiscale method (HMM)~\cite{Weinan2005, Niyonzima2013}
or the multilevel finite element method (FE$^2$)~\cite{Schroder2014},
the effective material response unavailable at the macroscopic scale
is computed during the simulation
by solving a mesoscopic-scale problem on a representative volume element
and homogenizing its solution.
In the present context,
the mesoscopic-scale model can be formulated
as a magneto-quasistatic FE simulation
solved on the cross-section of a lamination
and incorporating the EB hysteresis model.
However, this approach remains too computationally demanding
for routine machine design for three main reasons:
i) The mesoscopic-scale model must be queried
in each macroscopic finite element,
at each time step and at each nonlinear iteration.
ii) The mesoscopic-scale model must additionally provide
the corresponding Jacobian, $\partial\vec{H}/\partial\vec{B}$ or $\partial\vec{B}/\partial\vec{H}$,
to construct the macroscopic tangent stiffness matrix required for a Newton-Raphson scheme.
If obtained by finite differences,
the mesoscopic-scale problem must be solved for the nominal state,
and for additional perturbed states,
further increasing the number of mesoscopic-scale model queries.
iii) Each mesoscopic-scale query itself requires evaluating the EB model
at every element of the mesoscopic mesh.

Advances in machine learning have led to a rapid expansion
of neural network (NN)-based methods dedicated to magnetic modeling.
Among other applications,
NNs are used to predict core losses \cite{solimene2024hybrid},
to predict hysteresis model parameters \cite{Aquino2024hysteresis, zhang2023magnetic},
or to predict $\vec B$ from $\vec H$ and inversely
\cite{guo2025dynamic, bertolini2026development,serrano2022neural,li2023predicting}.
This enthusiasm is further evidenced by
the emergence of dedicated international competitions
in the power electronics community,
such as the MagNet challenges,
which have been specifically designed to stimulate innovation
in the data-driven modeling of core losses~\cite{chen2024magnet}
and transient magnetic behavior~\cite{wang2025magnetx}.
Yet, despite an intense research activity,
and proven effectiveness in related fields
such as heterogeneous elasto-plastic materials~\cite{wu2020recurrent},
integrating an NN-based surrogate model to account for vector hysteresis and eddy currents
in 2D FE simulations of realistic electrical machines remains an open problem.

In a previous contribution \cite{purnode2024neural},
the relation required for magnetic vector-potential formulations
was approximated by a parametric homogenized constitutive law
$\tilde{\vec H}(\vec B,\dot{\vec B},p_k)$.
The parameters $p_k$ of this law were identified element-wise
through a three-step procedure:
i) A 2D macroscopic-scale anhysteretic FE simulation
was performed to compute characteristic periodic $\vec H(t)$ sequences.
ii) Corresponding $\vec{B}(t)$ responses were obtained
by feeding the resulting $\vec{H}(t)$ sequences
to the aforementioned mesoscopic-scale model
using the explicit variant of the forward EB model,
and homogenizing its solution.
iii) The parameters $p_k$ were then identified by fitting
the homogenized constitutive law $\tilde{\vec{H}}(\vec{B}, \dot{\vec{B}}, p_k)$
to the computed $(\vec{B}(t), \vec{H}(t))$ pairs.
This identification was carried out through NN training,
which minimizes the fitting error over all sequences simultaneously
and yields a smooth spatial distribution of the parameters $p_k$
across the macroscopic domain.
The resulting macroscopic 2D FE simulations efficiently
captured the combined effects of hysteresis and eddy currents
through the parametric constitutive law $\tilde{\vec{H}}(\vec{B}, \dot{\vec{B}}, p_k)$.
However, this approach is restricted to periodic steady-state operating conditions.
Furthermore, the fixed analytical structure of the parametric law
inherently limits the achievable accuracy,
and the resulting macroscopic FE simulations may converge slowly.

In the present work, rather than assuming a fixed parametric structure,
an effective constitutive relation $\tilde{\vec H}(\vec B)$ is directly learned
from precomputed data generated by the aforementioned mesoscopic-scale model.
A recurrent neural network (RNN) is used to naturally handle
the history-dependent nature of hysteresis,
lifting the periodicity restriction of the previous approach,
while the Jacobian $\partial\tilde{\vec{H}}/\partial\vec{B}$
is obtained exactly through automatic differentiation~\cite{Baydin2017Automatic}.
The improved approach is evaluated
by modeling two real-world applications.
It demonstrates excellent accuracy and robust convergence properties,
while maintaining computational times
comparable to those of conventional anhysteretic simulations.
In addition, the proposed NN-based model provides an estimation of its error,
which constitutes valuable information for assessing the reliability
of the computed solution.

The paper is organized as follows.
Section \ref{sec:tgtModel} first presents the mesoscopic-scale model
used to generate the training datasets and to assess the NN-based model accuracy.
Section \ref{sec:NN} then details the model architecture, training dataset and training procedure,
while Section \ref{sec:FEM} presents the conducted 2D FE simulations.
Finally, Section \ref{sec:conclusion} provides conclusions and perspectives for future work.

%% file: lam.tex
This section presents the detailed lamination model
used as reference throughout this paper.
It provides a homogenized description of the fields at the mesoscopic scale,
i.e., at the scale of the lamination thickness.
The model combines the explicit variant of the EB hysteresis model \cite{henrotte2006energy}
with a one-dimensional (1D) magnetodynamic FE formulation accounting for eddy currents.
The EB model is first introduced, followed by the associated 1D FE simulation.

\subsection{Quasi-Static Mesoscopic Model: The Energy-Based Hysteresis Model}
\label{sec:EBM}

\begin{figure}[!htbp]
	\centering
	\includegraphics{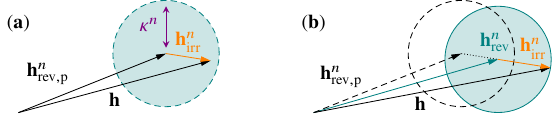}
	\caption{
		Illustration of (\ref{eq:update}).
		(\textbf{a}) $\vec h$ remains in the circle of radius $\kappa^n$ centered in $\vec h_{\text{rev},\text{p}}^n$,
		so that $\vec h_\textit{rev}^n$ remains unchanged.
		(\textbf{b}) $\vec h$ exits the circle,
		so that $\vec h_\text{rev}^n$ moves along the line linking $\vec h_{\text{rev},\text{p}}^n$ and $\vec h$,
		and stop at a distance $\kappa^n$ from $\vec h$.}
	\label{fig:EBM}
\end{figure}

The EB hysteresis model is grounded in thermodynamic principles;
its theoretical derivation is presented in detail in \cite{Jacques,franccois2013energy}.
Without expanding on this derivation,
the model captures the distributed nature of pinning strengths
through an ensemble of $N$ elementary cells.
Each cell $n$ decomposes the mesoscopic magnetic field $\vec h$
into a reversible component $\vec h_\text{rev}^n$ 
and an irreversible component $\vec h_\text{irr}^n$,
whose magnitude is constrained within a threshold $\kappa^n$.
As long as $|\vec h_\text{irr}^n|\le \kappa^n$, the cell behaves reversibly.
Once the threshold is exceeded, the cell transitions into irreversible behavior.
Among the update laws proposed to describe this transition,
we adopt the explicit variant, 
illustrated on Fig.~\ref{fig:EBM} and expressed as
\begin{equation}
	\vec h_\text{rev}^n = \begin{cases}
		\vec h_{\text{rev},\text{p}}^n & \text{if } |\vec h-\vec h_{\text{rev},\text{p}}^n| \le \kappa^n\text{,}\\
		\vec h - \kappa^n \frac{\vec h-\vec h_{\text{rev},\text{p}}^n}{|\vec h-\vec h_{\text{rev},\text{p}}^n|} & \text{otherwise,}
	\end{cases}
	\label{eq:update}
\end{equation}
where $\vec h_{\text{rev},\text{p}}^n$ denotes the previous reversible component $\vec h_\text{rev}^n$.
From the weighted sum of the reversible fields
$\vec h_\text{rev} = \sum_{n=1}^{N} w^n \vec h_\text{rev}^n$,
with the $w^n$ parameters summing up to 1,
the mesoscopic magnetic flux density is expressed as
\begin{equation}
	\vec b
	=
	\mu_0 \vec h
	+
	\left[
		\mathscr{L}\left(\frac{|\vec h_\text{rev}|}{h_a}\right)\frac{J_a}{|\vec h_\text{rev}|}
		+
		\mathscr{L}\left(\frac{|\vec h_\text{rev}|}{h_b}\right)\frac{J_b}{|\vec h_\text{rev}|}
	\right]
	\vec h_\text{rev},
	\label{eq:b}
\end{equation}
\noindent
where $\mu_0$ is the magnetic permeability of vacuum,
$\mathscr{L}\,(x)=\coth\,(x)-1/x$ denotes the Langevin function,
and $h_a$, $h_b$, $J_a$ and $J_b$ are parameters of the EB model.
These parameters, along with the $\kappa^n$ and $w^n$ parameters,
are determined from measurements,
and have to be updated whenever the material changes.
This work employs $N=11$ elementary cells
for the modeling of M235-35A non-oriented electrical steel sheets,
with the identified parameter values reported in \ref{sec:EBparams}.

\subsection{Dynamic Mesoscopic Model: Laminations with Eddy Current and Hysteresis}
\label{sec:lam}

\begin{figure}[!htbp]
	\centering
	\includegraphics{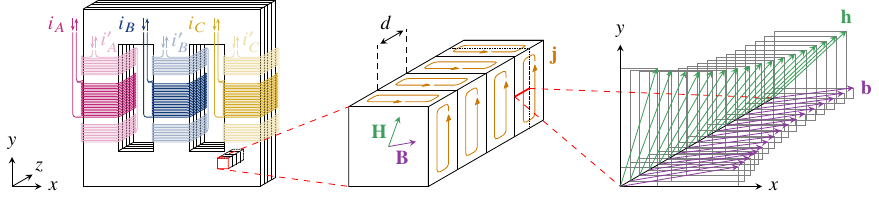}
	\caption{Definition of the domain of analysis of the 1D lamination model.}
	\label{fig:lam}
\end{figure}

The aforementioned EB model solely accounts for quasi-static hysteresis.
To also account for dynamic effects,
it is included into the FE simulation of a stack of ferromagnetic laminations.
Considering that all the laminations see the same imposed macroscopic magnetic field $\vec H$,
and neglecting edge effects,
simulating a single lamination is sufficient to describe the whole stack.
Furthermore, the mesoscopic magnetic field $\vec{h}$
and the mesoscopic magnetic flux density $\vec{b}$ inside a lamination
can reasonably be assumed parallel to the lamination plane
and to vary exclusively within the lamination depth.
Eventually, the symmetry with respect to the median plane of a lamination
is exploited to simulate solely half of a lamination thickness $d$ (see Fig.~\ref{fig:lam}).

Let $t$ denote time,
$\rho$ the electrical resistivity,
and $(x,y,z)$ a Cartesian coordinate system,
with $x$ and $y$ lying in the plane of the laminations
and $z$ in the stacking direction.
Furthermore, let $z=0$ correspond to the lamination surface
and $z=d/2$ to its center.
The 1D magneto-quasistatic formulation then reads:
Find the mesoscopic magnetic field $\vec{h}(z,t)=(h_x(z,t), h_y(z,t))$ such that
\begin{equation}
	\begin{cases}
		\partial_t\Big( b_x\left(\vec{h}\right) , h_x' \Big)_{\left[0,d/2\right]}
		+ \Big( \rho\,\partial_zh_x , \partial_zh_x' \Big)_{\left[0,d/2\right]}&=0\\
		\partial_t\Big( b_y\left(\vec{h}\right) , h_y' \Big)_{\left[0,d/2\right]}
		+ \Big(\rho\,\partial_zh_y , \partial_zh_y' \Big)_{\left[0,d/2\right]}&=0
	\end{cases}
	\label{eq:lam}
\end{equation}
holds for all test functions $h_x'$, $h_y'$, where the notation $(\cdot,\cdot)_D$
denotes the inner product of its arguments over bounded domain $D$.
The macroscopic excitation $\vec H(t)$ is imposed at the lamination surface,
i.e. $\vec{h}(0,t) = \vec H(t)$,
and the zero-current condition at the mid-plane leads to the Neumann boundary condition
$\partial_z\vec{h}(d/2,t)=0$.
The macroscopic magnetic flux density $\vec B(t)$ is retrieved by averaging
the mesoscopic magnetic flux density $\vec b(\vec h)=(b_x(\vec h),b_y(\vec h))$
over the half lamination thickness.
This formulation accounts for the eddy currents and the associated skin effect
in the thickness direction via the resistivity term $\rho\,\partial_z\vec{h}$,
and accounts for hysteresis via the $\vec b(\vec h)$ term,
solution of the EB model.

In practice, the model is implemented as a self-contained C language code,
solving the linear system, whose matrix is sparse,
using an efficient LU sparse algorithm.
The code is applied to a stack of M235-35A non-oriented electrical steel laminations,
whose domain of analysis $z \in \left[0,d/2\right]$
is discretized with 51 equidistant nodes.
Given that the maximum differential permeability is
$\mu_\text{max}=\lim_{h\rightarrow0^+}(\mu_0+\mathscr{L}(h/h_a)\,J_a/h+\mathscr{L}(h/h_b)\,J_b/h)\approx25.5$~mH/m,
obtained from (\ref{eq:b}),
the resulting mesh size $d/100=3.5~\mu$m ensures that the skin depth
$\delta = \sqrt{\rho/(\pi f \mu_\mathrm{max})}$
is resolved by at least five nodes up to approximately 44~kHz.
Thus, skin-effect phenomena are accurately captured over the frequency range considered in this work (see Section \ref{sec:data}).
The temporal discretization is chosen sufficiently fine to avoid nonlinear iterations,
and the time derivatives are discretized with the implicit Euler scheme.
Finally, the cells of the EB model are initialized
assuming an initially demagnetized material subjected
to a slowly increasing excitation field until reaching $\vec H(t=0)$,
such that no dynamic effects are initially present and $\vec B(\vec H(t=0))$
lies on the virgin magnetization curve.

%% file: NN.tex
The choice of the NN architecture is of predominant importance.
Since hysteresis inherently depends on the material history,
the NN model must operate on time-sequence data.
Some authors process the entire sequence at once using,
for example, neural operators \cite{2025_Chandra_A}.
In FE simulations, however, the input sequence length
evolves continuously during time integration,
which makes whole-sequence architectures poorly suited
for direct integration into incremental solvers.
Another approach consists in using a fixed-length sliding window
that captures only the most recent history
\cite{2023_Quondam,2024_Stella,2024_Vuokila}.
Nevertheless, such a fixed-length window may omit parts of the history
that significantly influence the response,
thereby reducing accuracy.
Alternatively, attention-based architectures \cite{2017_Vaswani}
have shown high accuracy on sequence-to-sequence tasks \cite{khan2022generalizable},
but the computational overhead becomes significant for long sequences.
In this contribution,
an RNN architecture is preferred.
Indeed, RNNs are well suited to handle hysteresis in FE simulations:
(i) they maintain and update state variables
similarly as the EB model maintains and updates the reversible fields $\vec h_\text{rev}^n$;
(ii) they are fast to infer a single time step,
making them efficient within incremental solvers;
and (iii) they achieve high accuracy \cite{2023_Li, 2025_Li, denis2026acceleration}.

This section discusses the implementation of the NN-based model
that will be integrated into the FE simulations of Section \ref{sec:FEM},
and details its training process.
At first, Section \ref{sec:GRU} presents the general model architecture
and Section \ref{sec:data} details the datasets used for training.
Section \ref{sec:MSSE} later
demonstrates that the model cannot be trained naively to minimize the mean squared error,
since saturated fields will dominate the training
to the detriment of lower fields where the magnetic behavior is more intricate.
Section \ref{sec:MSSE} therefore introduces a scaled loss,
which is further modified in Section \ref{sec:NNLL} to assess the model confidence.
In Section \ref{sec:robustness}, the model is upgraded to ensure correct prediction
even in extreme saturation, which lies out of the training range.
Section \ref{sec:resample} later describes how the training sequences,
initially defined on constant time steps,
are resampled to include variable time steps,
thereby enabling adaptive time-stepping strategies in the FE simulations.
Finally, Section~\ref{sec:train} discusses the hyperparameter tuning,
while Section \ref{sec:perf} assesses the model performance,
prior to any FE coupling.

\subsection{Baseline Architecture}
\label{sec:GRU}

To process temporal sequences, RNNs ingest a single time step at a time.
At each time step~$k$, the network updates a hidden state~$S^k$
based on the current input~$X^k$ and the previous hidden state~$S^{k-1}$.
In its simplest form, this update is given by $S^{k}=\phi(W_x\,X^k+W_s\,S^{k-1}+b)$,
where $\phi(\cdot)$ is a nonlinear activation function,
$W_x$ and $W_s$ are weight matrices, $b$ is a bias term,
and the output~$Y^k$ is computed directly from the hidden state~$S^k$.
This basic formulation however suffers from
the vanishing- and exploding-gradient problems,
which impair its ability to learn long-range dependencies.
To address this issue, advanced recurrent units have been developed,
the most prominent being the Long Short-Term Memory (LSTM) \cite{1997_Hochreiter}
and the Gated Recurrent Unit (GRU) \cite{2014_cho}.
The GRU is adopted here,
as it achieves accuracy comparable to LSTMs
with fewer parameters and a simpler gating structure,
making it both easier to train and more efficient to deploy.

\begin{figure}[ht]
	\centering
	\includegraphics[]{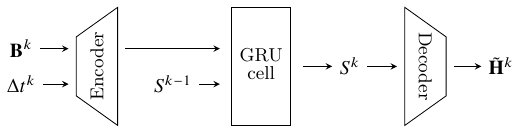}
	\caption{
		Baseline model architecture, enhanced in Sections \ref{sec:NNLL} and \ref{sec:robustness}.}
	\label{fig:GRU}
\end{figure}

Without going into the details of the GRU cell,
Fig.~\ref{fig:GRU} presents the basic model architecture,
which is further enhanced in Sections~ \ref{sec:NNLL} and \ref{sec:robustness}.
At time step $k$,
the model takes as input the $x$- and $y$-components of the magnetic flux density $\vec B^k$,
together with the time increment $\Delta t^k$.
The latter may vary between successive iterations
to enable temporal refinement during FE simulations.
Both $\vec B^k$ and $\Delta t^k$ are scaled, concatenated,
and passed through an encoder block implemented as a multilayer perceptron,
which progressively increases the feature dimensionality
to match the GRU hidden-state size.
The GRU cell is followed by a decoder block,
also implemented as a multilayer perceptron,
which maps the hidden state back to the $x$- and $y$-components of $\tilde{\vec H}^k$,
that is the predicted magnetic field,
intended to approximate the target magnetic field $\vec H^k$.
Rather than initializing the hidden state to zero,
it is initialized by a trainable vector,
plus the output of a shallow network taking $\vec B^0$ and $\Delta t^0$ as inputs.
This offloads the initial magnetization response,
which is history-independent and lies on the virgin curve (see Section \ref{sec:lam}),
to a dedicated subnetwork.
The GRU cell therefore starts with a meaningful hidden state,
and can devote its full capacity to learning history-dependent evolution,
reducing the number of parameters that would otherwise be required.

\subsection{Generation of the Training Datasets}
\label{sec:data}

In order to propose an NN-based surrogate model
readily usable in a wide range of applications,
the training dataset should be populated with a broad variety of field profiles.
Since using data from simulations or measurements on existing devices
may bias the model toward specific applications,
the training $\vec H$ sequences are generated artificially
to span the diversity of waveforms encountered in electrical machines.
The corresponding $\vec B$ sequences are later obtained by feeding them
to the lamination model of Section~\ref{sec:tgtModel}.
As an initial strategy,
the training $\vec H$ sequences can be expressed
as a sum of harmonics in which each component is defined
by two orthogonal contributions with random amplitudes and phases:

\begin{equation}
	\vec H(t)=\sum_{h\in\mathcal{H}}\frac{\alpha_h H^\star}{\sqrt{h}}
	\left[
	\cos(\beta_h)\sin(2\pi fht+\varphi_h^x)\,\hat{\vec x}
	+\sin(\beta_h)\sin(2\pi fht+\varphi_h^y)\,\hat{\vec y}
	\right].
	\label{eq:dataBasic}
\end{equation}

\noindent
In this expression,
the set $\mathcal{H}$ is generated by first randomly selecting
the maximum harmonic $h_\text{max}\in[1,50]$,
later randomly selecting the number of harmonics $|\mathcal{H}|\in[1,h_\text{max}]$,
and finally randomly selecting the remaining harmonic indices without replacement
in $\{1,\ldots,h_\text{max}-1\}$.
The characteristic amplitude $H^\star$ is drawn
from a log-uniform distribution over $[10~\mathrm{A/m}, 100~\mathrm{kA/m}]$,
while the fundamental frequency $f$ is uniformly sampled between 0.1~Hz and 1~kHz,
with harmonic $h$ having frequency $h\,f$.
The selected ranges for $H^\star$ and $h_\text{max}$
are justified in Sections~\ref{sec:robustness} and~\ref{sec:resample}, respectively,
where the associated anhysteretic extension and resampling strategy are introduced.
The coefficients $\alpha_h \in [0,1]$ are also uniformly sampled and,
together with the decay factor $1/\sqrt{h}$,
modulate the relative amplitude of each harmonic,
thereby controlling the harmonic distribution and enabling the generation
of both smooth and highly distorted waveforms.
The parameters $\beta_h \in [0,2\pi]$ in turn govern
how each harmonic contribution is shared
between the orthogonal directions $\hat{\vec x}$ and $\hat{\vec y}$,
thus defining its orientation in the plane.
Eventually, the phase terms $\varphi_h^x$ and $\varphi_h^y \in [0,2\pi]$
independently shift the temporal evolution of each directional contribution,
while their difference determines the trajectory of each harmonic component in the plane:
identical phases lead to unidirectional oscillations,
whereas a non-zero phase difference produces a rotating component,
resulting in circular or elliptical trajectories.

Up to now, (\ref{eq:dataBasic}) describes a broad class of excitations,
encompassing alternating, rotating, and distorted magnetic field trajectories.
However, it poorly accounts for several features commonly encountered in electrical machines:
i)~purely unidirectional fields,
ii)~DC-biased fields,
iii)~pulse-like fields,
iv)~and ramp-up fields.
To improve the representation of these excitation types,
several independent modifications are introduced.
Each modification is activated with a probability of 50~\%,
avoiding both under-representation of the corresponding waveform characteristics and excessive bias toward them.

\paragraph{i) Unidirectional Fields}

Without specific constraints on the parameters
$\beta_h$, $\varphi_h^x$, and $\varphi_h^y$,
(\ref{eq:dataBasic}) almost never generates purely unidirectional fields.
Such fields arise only when all $\beta_h$ are identical and when
$\varphi_h^x$ and $\varphi_h^y$ are either equal
or differ by an integer multiple of $\pi$
for all $h \in \mathcal{H}$.
However, unidirectional fields are commonly encountered in electrical machines.
To ensure their adequate representation in the training dataset,
all $\beta_h$ are assigned the same value with a probability of 50~\%.
Independently, the phase terms are generated such that
$\varphi_h^x = \varphi_h^y$ for all $h \in \mathcal{H}$,
again with a probability of 50~\%.
Since these two conditions are imposed independently,
approximately 25~\% of the generated sequences are purely unidirectional.

\paragraph{ii) DC-biased Fields}

Among other environments,
DC-biased fields are encountered in the rotor of synchronous machines,
in the stator of DC machines and in the presence of permanent magnets.
With a probability of 50~\%,
(\ref{eq:dataBasic}) is therefore augmented by a DC component
\begin{equation}
	\vec H_\text{dc}=H_\text{dc}\,\Big(\cos\left(\beta_\text{dc}\right)\,\hat{\vec x}+\sin\left(\beta_\text{dc}\right)\,\hat{\vec y}\Big)
\end{equation}
\noindent
where both the amplitude $H_\text{dc}\in[0,H^\star]$
and direction $\beta_\text{dc}\in[0,2\pi]$
are uniformly sampled.

\paragraph{iii) Pulse-Like Fields}

In rotating electrical machines,
the slotted geometry of the stator and rotor often causes
the air-gap reluctance to vary periodically.
As a result, the field near the air-gap takes a pulse-like shape:
high when rotor and stator teeth are facing each other,
and significantly reduced when a tooth faces a slot.
In principle, (\ref{eq:dataBasic}) can build such pulse-like fields,
but it is unlikely.
To ensure adequate coverage of this field shape in the dataset,
a pulse-like modulation is applied with a probability of 50~\%.
When applied, the $x$ and $y$ components of both (\ref{eq:dataBasic}) and $\vec H_\text{dc}$
are scaled by $\alpha_p^x + (1-\alpha_p^x)\,p(t)$
and $\alpha_p^y + (1-\alpha_p^y)\,p(t)$, respectively,
where $\alpha_p^x$ and $\alpha_p^y \in [0,1]$ are uniformly sampled, and
\begin{equation}
	p(t)=\sin(2\pi fh_pt+\varphi_p)^{e_p}
\end{equation}
\noindent
is a pulse-like function.
The number of pulses per period $h_p\in[1,20]$ and the phase $\varphi_p\in[0,2\pi]$
are both drawn from uniform distributions. 
The upper bound on $h_p$ was chosen to ensure
the resulting pulses are sufficiently resolved,
while still allowing variability within each pulse
owing to the harmonic content of (\ref{eq:dataBasic}).
On the other hand, the exponent $e_p$,
sampled as a random integer between 3 and 20,
controls the sharpness and symmetry of the pulses.
When $e_p$ is odd, $p(t)$ preserves the sign of the sine term,
resulting in alternating positive and negative pulses,
as can be observed in the stator teeth.
Conversely, when $e_p$ is even,
the function remains non-negative,
producing pulses of constant sign,
as is typical in the rotor near the air gap,
a behavior that will be illustrated in Fig.~\ref{fig:curveRotorAirgap}
at the end of Section~\ref{sec:4e}.

\paragraph{iv) Ramp-up Transients}

The magnetic excitation can progressively build up from an initially demagnetized state.
To account for such excitations,
the generated field is multiplied, with a probability of 50~\%, by the ramp-up function
\begin{equation}
	r(t)=1-\exp(-\gamma\pi ft)
\end{equation}
where $\gamma$ is drawn from a log-uniform distribution over $[1, 10]$.
This parameter controls the rise speed of the excitation,
ranging from smooth progressive growth to
rapid establishment of the underlying waveform.

\vspace{0.35cm}

From this procedure, $100\,000$ $\vec H$ sequences are generated
and fed to the lamination model to form the training dataset.
As will be shown in Section~\ref{sec:perf},
this dataset size provides sufficient accuracy.
Smaller datasets, on the other hand, lead to a degradation in model performance.
While the lamination model is evaluated with a fine temporal discretization,
as discussed in Section~\ref{sec:lam},
the resulting sequences are downsampled to 500 uniform time steps
spanning half of the fundamental period $T=1/f$.
Although the second half is not strictly redundant ---
owing to phase differences, pulse modulation, and ramp-up transients ---
it contains largely similar information.
Furthermore, restricting the sequence length speeds up the RNN training
while reducing the risk of vanishing or exploding gradients.
Validation and test datasets are generated in the same manner, each comprising $10\,000$ sequences.
Overall, the dataset generation process requires approximately four days
on a single CPU core of an HPE ProLiant XL225n compute node from the Lucia cluster.
The workload is, however, embarrassingly parallel
and can therefore be completed in a much shorter wall-clock time
when distributed across multiple CPU cores.

\subsection{Permeability-Based Error Scaling}
\label{sec:MSSE}

A common loss for regression tasks is the Mean Squared Error (MSE).
Restraining its definition to the time step~$k$ of a single sequence,
it is given by

\begin{equation}
	\mathcal{L}_\text{MSE}
	=
	\left|\vec H^k-\tilde{\vec H}^k\right|^2.
\end{equation}

\begin{figure}[ht]
	\centering
	\includegraphics[]{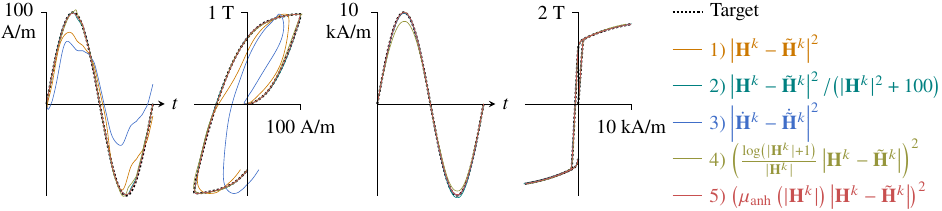}
	\caption{Given a low-field (left) and a high-field (right)
		sinusoidal $\vec H$ curves oscillating at 200~Hz with 500 time steps per period,
		and the corresponding $\vec B$ sequences returned by the lamination model,
		comparison between the $\tilde{\vec{H}}(\vec B)$ sequences
		returned by models trained to minimize different losses:
		1) the MSE,
		2) a relative loss,
		3) a loss on the time derivative,
		4) a loss using a logarithmic scaling,
		and 5) the proposed MSSE.
		All models were trained on the same datasets with uniformly spaced time steps
		and using identical numbers of layers, hidden units, and activation functions,
		as adopted for the final architecture in Section \ref{sec:train}.
		A grid search over the learning rate and batch size was performed for each loss individually.
		Among the losses, the proposed MSSE loss shows the best performance
		in both low- and high-field regions.
	}
	\label{fig:MSSE}
\end{figure}

\noindent
Poor results are however obtained using this loss as is.
Indeed, in the saturated regime, the magnetic response can vary over tens of kA/m or more,
so even a small relative error produces a large absolute contribution to the loss.
Conversely, low-field regions (where the material behavior is most sensitive and non-linear)
contribute comparatively little to the total error, leading to poor accuracy in this regime.
Common strategies to improve accuracy across a wide range of output values include:
(i) using a relative loss,
e.g. $|\vec H^k-\tilde{\vec H}^k|^2/(|\vec H^k|^2+\epsilonup)$,
with an $\epsilonup$ added in the denominator term
to control the trade-off between accuracy at low and high fields;
(ii) defining the loss on the time derivative, although it may suffer the same issues as the MSE,
and may cause the predicted trajectories to drift progressively over time; and
(iii) applying a logarithmic scaling, which improves overall accuracy but does not fully resolve the imbalance between low- and high-field regions.
To overcome these limitations,
this paper introduces a loss that conjointly scales
the predicted field $\tilde{\vec{H}}^k$ and the target field $\vec H^k$
by the anhysteretic magnetic permeability $\mu_\text{anh}(|\vec H^k|)$.
This scaling essentially expresses the error in terms of a magnetic flux density,
naturally reducing the dominance of high saturated fields.
The resulting loss, referred to as the Mean Squared Scaled Error (MSSE),
is specifically designed for applications involving magnetic saturation
and writes

\begin{equation}
	\mathcal{L}_\text{MSSE}
	=
	\left(\mu_\text{anh}\left(|\vec H^k|\right)\left|\vec{H}^k-\tilde{\vec{H}}^k\right|\right)^2,
	\label{eq:MSSE}
\end{equation}

\noindent
with

\begin{equation}
	\mu_\text{anh}\left(|\vec H^k|\right)
	= \mu_0
	+ \mathscr{L}\left(\frac{|\vec H^k|}{h_a}\right)\frac{J_a}{|\vec H^k|}
	+ \mathscr{L}\left(\frac{|\vec H^k|}{h_b}\right)\frac{J_b}{|\vec H^k|}
	\label{eq:mu},
\end{equation}

\noindent
which directly stems from (\ref{eq:b}).
Fig.~\ref{fig:MSSE} demonstrates the suitability of (\ref{eq:MSSE}),
by comparing the predictions made on two simple sequences
after having trained the model of Fig.~\ref{fig:GRU} on the aforementioned losses.

\subsection{Learning the Prediction Confidence}
\label{sec:NNLL}

Beyond predicting the material response itself,
it is valuable for the NN to also indicate how much confidence
it places in each of its own predictions.
Nix and Weigend \cite{1994_Nix} proposed in 1994 a loss function
that enables an NN to learn both the mean and standard deviation
of a normal probability distribution,
originally to capture the uncertainty inherent to data collection.
Although this source of uncertainty is absent here --
since the lamination model used to generate the training dataset is deterministic
(see Sections~\ref{sec:tgtModel} and \ref{sec:data}) --
a similar loss can be used to make the NN predict an expected error.
Evaluating the NN previously-trained with the MSSE loss
on the whole validation dataset shows that the error
$\mu_\text{anh}(|\vec H^k|)\,|\vec H^k-\tilde{\vec H}^k|$ is exponentially distributed.
Following the steps presented in~\citep[Section~5.2]{2023_Prince},
the loss is therefore derived by taking the negative log-likelihood
of an exponential distribution with scale parameter $\varepsilon$.
This yields the exponential negative log-likelihood (ENLL) loss,
adopted as the final training loss and
expressed here for a single time step $k$:
\begin{equation}
	\mathcal{L}_\text{ENLL}
	=\log\left(\varepsilon^k\right)
	+\frac{\mu_\text{anh}(|\vec H^k|)\,|\vec H^k-\tilde{\vec{H}}^k|}{\varepsilon^k}.
\end{equation}
\noindent
The two terms in the ENLL loss play complementary roles:
the logarithmic term penalizes overly large values of $\varepsilon^k$,
whereas the second term penalizes values that underestimate the prediction error.
Their balance encourages $\varepsilon^k$ to match the actual prediction error.
Consequently, the scale parameter $\varepsilon^k$ acts as a predicted error,
expressed in tesla,
which varies from sequence to sequence and from time step to time step.
At evaluation, $\varepsilon^k$ is expected to grow
when the model is given sequences it poorly handled during training.
In practice, $\varepsilon^k$ is also expected to grow
on sequences lying outside of the training domain,
but this behavior remains a tendency rather than a certainty,
as the NN response to unseen inputs cannot itself be guaranteed.

To ensure stable training,
$\varepsilon^k$ must remain strictly positive.
In practice, a minimum value of $\varepsilon_\text{min}=1$~mT is enforced
using a softplus-like activation function.

\begin{equation}
	\varepsilon^k = \frac{1}{\varkappa}\ln\left[1+\exp\bigg(\varkappa\left(\varsigma o^k+1-\varepsilon_\text{min}\right)\bigg)\right] + \varepsilon_\text{min},
\end{equation}

\noindent
where $o^k$ denotes the raw output of the decoder,
$\varsigma=0.1$ controls the slope of the activation function,
and $\varkappa=5$ its sharpness.

\subsection{Robustness at High Fields}
\label{sec:robustness}

The model is trained on $\vec H$ sequences
with maximum amplitudes of roughly $100$~kA/m (see Section \ref{sec:data}),
which, via (\ref{eq:mu}), correspond to anhysteretic amplitudes
$|\vec B|=\mu_\text{anh}(|\vec H|)\,|\vec H|$
of approximately $2.05$~T.
The model is therefore expected to provide reliable predictions within this range.
Predictions for $\vec B$ amplitudes exceeding $2.05$~T, however,
lie outside the training domain and their accuracy cannot be guaranteed.
While exceeding $2.05$~T is uncommon in practice,
excursions to higher field values may occur
during the Newton–Raphson iterations
of the 2D macroscopic-scale FE simulations (see Section \ref{sec:FEM}).
Therefore, the model must remain well-behaved even when $|\vec B|$ exceeds 2.05~T.

\begin{figure}[ht]
	\centering
	\includegraphics[]{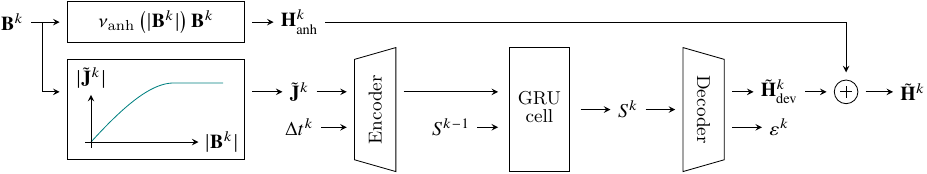}
	\caption{Enhanced model architecture.}
	\label{fig:NN}
\end{figure}

Beyond $100$~kA/m, ferromagnetic materials are fully saturated.
In this regime, the material response follows a simple anhysteretic law, such as (\ref{eq:mu}).
A practical strategy is thus to detect when the model operates outside its training domain,
and to replace it with an anhysteretic law.
However, this approach may introduce issues near the switching point, as it may create a discontinuity.
The proposed solution relies on two points:
i) having the neural network predict only the deviation from an anhysteretic curve; and
ii) clipping the amplitude of the input $\vec B^k$,
ensuring that the model never receives data outside its training domain.
The resulting enhanced model architecture is illustrated in Fig.~\ref{fig:NN}.

\paragraph{i) Residual formulation}

The predicted field $\tilde{\vec H}^k$ is now written as the sum of
an anhysteretic contribution $\vec H_\text{anh}^k$
and a hysteretic deviation $\tilde{\vec{H}}_\text{dev}^k$, output of the decoder.
The term $\vec H_\text{anh}^k$ is computed as $\nu_\text{anh}(|\vec B^k|)\,\vec B^k$,
where $\nu_\text{anh}(|\vec B^k|)=1/\mu_\text{anh}(|\vec B^k|)$ expresses a magnetic reluctivity,
and  is implemented as a piecewise cubic Hermite interpolating polynomial (PCHIP).
By inverting (\ref{eq:mu}), it is defined on a uniform partition of $|\vec B|^2$,
each interval having a width $\Delta B^2$.
The index of the appropriate spline segment
is therefore directly given by $\lfloor (B_x^2+B_y^2)/\Delta B^2\rfloor$.
If the PCHIP was defined on $\Delta B$ intervals,
an additional square root would be required.
Alternatively, if it was defined on $\Delta H$ intervals,
finding the corresponding spline index would require a lookup operation.
For $|\vec B^k|\ge3$~T, the PCHIP is replaced
by the analytical expression $[(|\vec B^k|-B_s)/\mu_0+H_s]/|\vec B^k|$,
where the parameters $B_s$ and $H_s$ are chosen to guarantee
$\nu_\text{anh}(|\vec B^k|)$ is continuous at $3$~T.

\paragraph{ii) Input Clipping}

When the material is fully saturated,
$\vec{H}^k(\vec B^k)$ follows the anhysteretic law,
so that $\vec H_\text{anh}^k(\vec B^k)$ alone captures the behavior of $\tilde{\vec{H}}^k(\vec B^k)$ and $\tilde{\vec{H}}_\text{dev}^k(\vec B^k)$ remains constant.
In this situation, the magnitude of the input $\vec B^k$ therefore
has no impact on the prediction and can be clipped without impacting the model accuracy.
Equivalently, the NN can be given an approximate magnetic polarization
$\tilde{\vec J}^k=[1-\mu_0\,\nu_\text{anh}(|\vec B^k|)]\,\vec B^k$
as input, since it is naturally bounded by $J_\text{max}=J_a+J_b\approx1.95$~T (see Section \ref{sec:EBM}).
With a training dataset populated with $\vec B$ sequences whose maximum amplitude reaches
$B_\text{max}\approx 2.05$~T,
the NN is trained on magnetic polarizations up to
$[1-\mu_0\,\nu_\text{anh}(B_\text{max})]\,B_\text{max}\approx 1.92$~T,
which is slightly smaller, yet sufficiently close to $J_\text{max}=1.95$~T.
Note that extending the training dataset beyond $100$~kA/m is therefore unnecessary.
It would also be counterproductive,
as it would increase the cost of data generation and training,
despite the saturated regime being fully described by $\vec H_\text{anh}^k$.

\vspace{0.35cm}

In addition to ensuring robustness at high fields,
the residual formulation offers further advantages
related to the learning problem itself.
By constraining the network to predict only the hysteretic deviation $\tilde{\vec H}_\text{dev}^k$
from the known anhysteretic baseline $\vec H_\text{anh}^k$,
the target function is significantly smoother and smaller in magnitude
than the full material response.
This simplification reduces the complexity of the function the NN must approximate,
allowing good accuracy to be achieved with a more compact architecture.
Furthermore, since the dominant, well-understood part of the behavior is handled analytically,
the NN is implicitly guided toward physically meaningful predictions,
which tends to reduce the risk of spurious oscillations.

\subsection{Sequence Resampling for Adaptive Time-Stepping}
\label{sec:resample}

At each training iteration,
a batch of sequences is drawn at random from the $100\,000$ sequences
composing the training dataset.
To encourage accurate predictions across different sequence lengths,
these sequences are randomly truncated,
resulting in sub-sequences of lengths between 10 and 500 uniformly-spaced time steps.
These sub-sequences are then resampled
to expose the model to non-uniform temporal discretizations,
making the trained model compatible with adaptive time-stepping schemes.

The complete resampling algorithm is presented in \ref{sec:resamplingAlg}.
In brief, the algorithm first selects a random subset of points to retain from the source sub-sequence,
while enforcing a maximum gap $\bar g = 5$ between consecutive retained points.
Since the highest considered harmonic ($50^\text{th}$ harmonic)
is initially sampled by 20 points per period (see Section~\ref{sec:data}),
enforcing $\bar{g} = 5$ guarantees that at least 4 points per period are retained after resampling, keeping it sufficiently resolved.
The algorithm then inserts up to $n_\text{ins}$ linearly interpolated points
according to a skewed random distribution,
allowing the interpolated points to be either smoothly spread
across the source sub-sequence or concentrated within localized regions.
In the present work, $n_\text{ins}$ is chosen equal to the length of the source sub-sequence,
allowing substantial variation in the time steps
while keeping the resampled sequence length comparable to that of the source sub-sequence.
This limits training time and mitigates vanishing and exploding gradient issues.
A minimum normalized spacing $\delta_\text{min} = 0.05$ is also enforced
between consecutive output points within each segment,
corresponding to 5~\% of the source-sequence time step $\Delta t$.
This minimum spacing prevents the insertion of nearly coincident points
that could cause numerical instabilities during training.
With this spacing, up to $1/\delta_\text{min}-1=19$ points can be inserted within a segment.
However, the algorithm limits the number of inserted points per segment to $\bar m = 10$,
in order to leave sufficient room for the remaining points
to be distributed with varying, non-degenerate time steps.
Moreover, during training, the loss is evaluated only at the retained source points
to prevent biasing the model toward the interpolation scheme.
Note that Neural Ordinary Differential Equations (Neural ODEs)
could inherently handle variable time steps,
as Neural ODEs learn the system dynamics and integrates it over the requested time interval,
rather than directly predicting the system output at a given time.
In the context of constitutive surrogate models for FE simulations,
this approach may however be inappropriate,
since it may require multiple NN evaluations to perform a single nonlinear iteration,
thereby increasing the computational cost.

\subsection{Hyperparameter Tuning and Training}
\label{sec:train}

\begin{figure}[ht]
	\centering
	\includegraphics[scale=1.0]{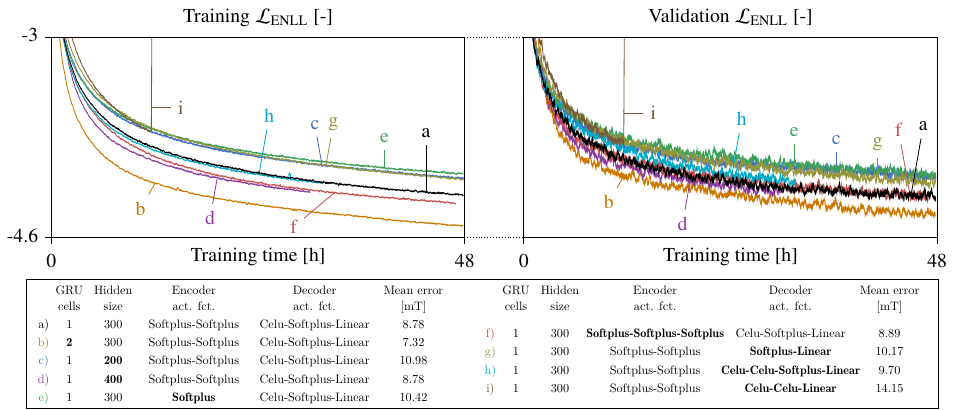}
	\caption{Training, validation, and test metrics for different architectural variants,
		each differing from the selected architecture a) by a single parameter.
		All models were trained using the Adam optimizer,
		a learning rate of $2\cdot10^{-4}$ and a batch size of 128.
		While the ENLL loss is used for training,
		the error $\mu_\text{anh}(|\vec H|)\,|\vec H-\tilde{\vec H}|$ is used on the test dataset.
		For readability, a moving-average operator was applied
		on the training and validation curves,
		using a window of 30.}
	\label{fig:train}
\end{figure}

Fig.~\ref{fig:train} compares the training, validation, and test metrics
obtained for different architectural variants,
and justifies the hereunder selected architecture.
Preliminary experiments revealed that using a single GRU cell
is sufficient to achieve high quality predictions.
Slightly better results are observed by stacking two GRU cells but,
since it doubles the evaluation time,
a single GRU cell is preferred.
This GRU cell uses a hidden size of 300,
chosen as a compromise between accuracy and computational cost.
The encoder and decoder use respectively two and three layers,
as deeper architectures did not improve performance significantly,
whereas shallower configurations were found less accurate.
The decoder's final layer uses a linear activation function
to produce continuous and unbounded $\tilde{\vec H}_\text{dev}^k$ values,
whereas the activation functions used in the encoder
and in the first two decoder layers
were determined through a grid search.
To guarantee a smooth Jacobian $\partial \vec H/\partial \vec B$
and thus well-behaved Newton–Raphson iterations,
the activation function search was restricted
to continuously differentiable candidates:
tanh, softplus, and CELU.
Although different activation function combinations produced similar final results,
some exhibited instabilities during training.
The final architecture uses softplus activations in both encoder layers,
and CELU and softplus activations in the first and second decoder layers, respectively.

The Adam optimizer was employed throughout the optimization procedure,
while the learning rate and batch size were progressively adjusted,
resulting in a final learning rate of $2\cdot10^{-4}$ for a batch size of 128.
Overall, the NN-based model contains approximately 670\,000 parameters,
corresponding to a storage size of 2.8~MB.
It was trained two days on a single NVIDIA A100 GPU from the Lucia cluster
using Pytorch \cite{pytorch}.
Although training was performed in single precision (float32),
inference is carried out in double precision (float64),
as the round-off errors inherent to single-precision arithmetic
were observed to prevent the FE simulations of Section~\ref{sec:FEM}
from reaching sufficiently small residuals.
Importantly, no additional training or fine-tuning was performed;
all subsequent results reported in this work were obtained using this single trained model.

\subsection{Standalone Performance}
\label{sec:perf}

\begin{figure}[ht]
	\centering
	\includegraphics[scale=1.0]{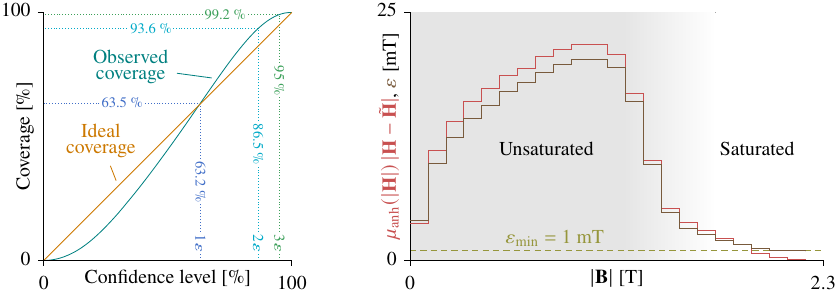}
	\caption{Comparison between the observed error 
			$\mu_\text{anh}(|\vec H|)\,|\vec H-\tilde{\vec H}|$
			and the predicted error $\varepsilon$ on the test dataset.
	Left: Coverage curve comparing the observed and ideal coverages,
	showing good overall calibration with slight overconfidence at low confidence levels,
	and slight underconfidence at high confidence levels.
	Right: Mean observed error and mean predicted error
	in bins of the magnetic flux density magnitude $|\vec B|$.
	}
	\label{fig:cov}
\end{figure}

To assess the model performance,
the error $\mu_\text{anh}(|\vec H|)\,|\vec H-\tilde{\vec H}|$ is evaluated
on the test dataset described in Section~\ref{sec:data},
yielding a low average value of 8.78~mT.
This error is defined as the norm of the difference between
the target $\vec H$ and predicted $\tilde{\vec H}$ magnetic field vectors,
and expressed in T through the scaling factor $\mu_\text{anh}(|\vec H|)$,
allowing a direct comparison with the predicted error $\varepsilon$.
Assuming the error follows an exponential distribution,
the expected fractions of predictions whose observed error lies within the
$1\varepsilon$, $2\varepsilon$, and $3\varepsilon$
predicted error intervals
are 63.2~\%, 86.5~\% and 95~\%, respectively.
In practice, the observed fractions on the test dataset are
63.5~\%, 93.6~\%, and 99.2~\%.
The coverage curve shown on the left of Fig.~\ref{fig:cov}
extends this comparison to all confidence levels
by plotting the observed proportion of predictions
lying within the corresponding predicted error interval.
A perfectly calibrated model would yield a coverage curve
lying exactly on the diagonal,
where the observed coverage systematically equals the nominal confidence level.
Curves below the diagonal indicate overconfidence,
whereas curves above the diagonal correspond to conservative error estimates.
As shown in Fig.~\ref{fig:cov},
the model is slightly overconfident for small confidence levels,
where the predicted error intervals are somewhat too narrow,
and slightly conservative at higher confidence levels.
The right panel of Fig.~\ref{fig:cov} also depicts how
the mean observed error and mean predicted error
vary with the magnetic flux density magnitude $|\vec B|$.
In particular, both decrease as saturation is reached,
reflecting the fact that the magnetic response becomes essentially anhysteretic,
so that $\vec H_\text{anh}$ alone provides an accurate prediction.
In the strongly saturated regime,
the observed error drops below the imposed lower bound $\varepsilon_\text{min}$,
making the predicted error conservative.

\begin{figure}[!htbp]
	\centering
	\includegraphics[scale=1.0]{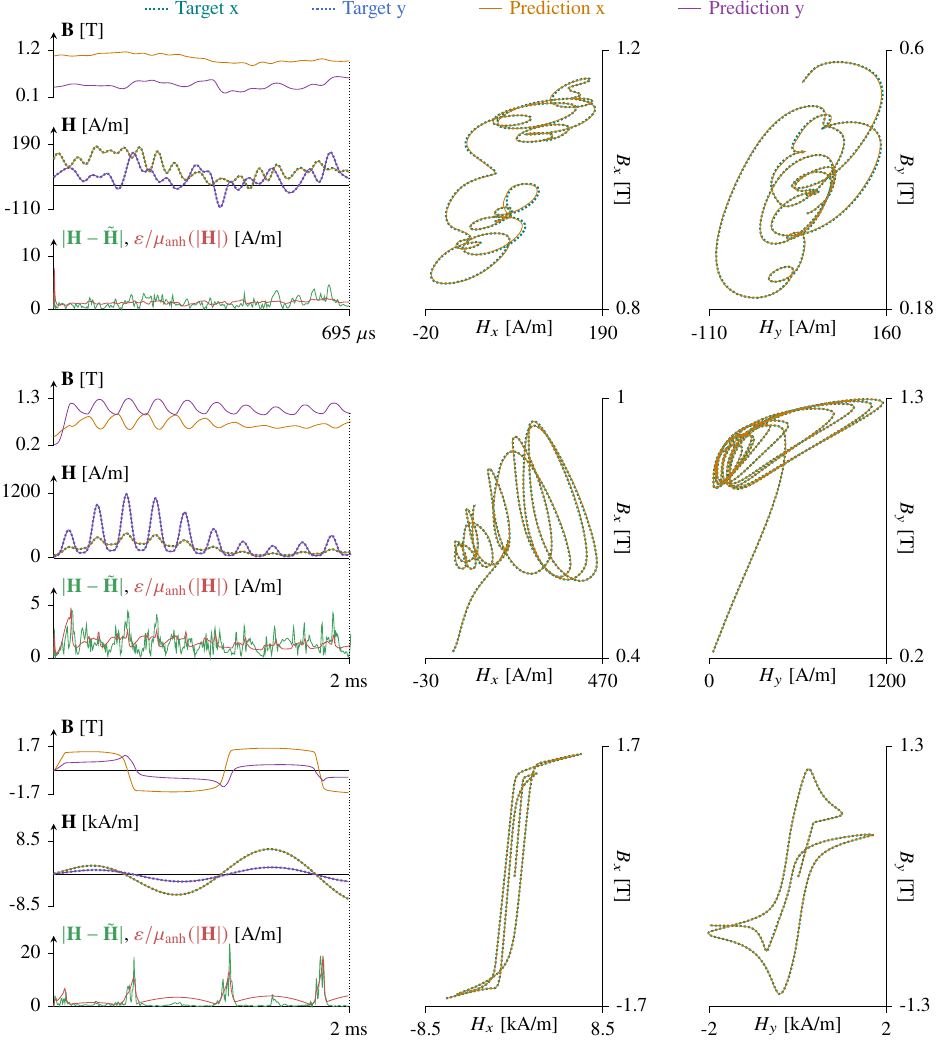}
	\caption{Comparison between three target $\vec H$ sequences from the test dataset,
		and the corresponding predicted $\tilde{\vec{H}}$ sequences.
		The mismatch $|\vec H-\tilde{\vec H}|$
		and the predicted error $\varepsilon/\mu_\text{anh}(|\vec H|)$
		show a good correlation.
	}
	\label{fig:testCurves}
\end{figure}

Fig.~\ref{fig:testCurves} compares three target $\vec H$ sequences
from the test dataset,
representative of low-, medium-, and high-field regimes,
with the corresponding predicted $\tilde{\vec{H}}$ sequences.
It also reports the observed error $|\vec H-\tilde{\vec H}|$
and the scaled predicted error $\varepsilon/\mu_\text{anh}(|\vec H|)$.
The figure demonstrates close agreement between
the target and predicted sequences
across the full range of field amplitudes,
including regions where small variations in $\vec B$
produce significantly amplified responses in $\vec H$.
Moreover, $\varepsilon/\mu_\text{anh}(|\vec H|)$ provides a good approximation
of the error $|\vec H-\tilde{\vec H}|$.

\begin{figure}[!htbp]
	\centering
	\includegraphics[]{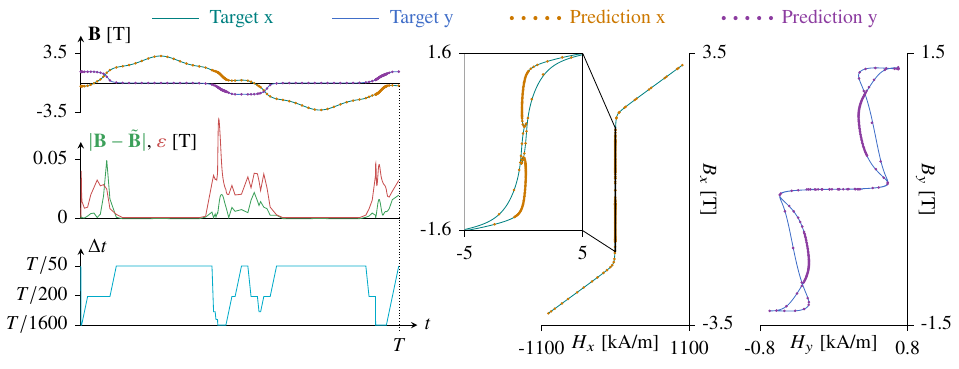}
	\caption{
		Inversion of the NN-based surrogate model via Newton-Raphson
		for a test $\vec{H}$ field spanning low- and high-field regimes,
		using an adaptive time-stepping strategy.
		}
	\label{fig:testDtVar}
\end{figure}

Similarly, Fig.~\ref{fig:testDtVar} compares the $\tilde{\vec{B}}$ field,
obtained by inverting the NN-based model
via Newton-Raphson for a given $\vec{H}$ field,
against the reference lamination solution $\vec{B}$.
The Jacobian $\partial \tilde{\vec{B}} / \partial \tilde{\vec{H}}$ required by the scheme
is obtained via automatic differentiation of the NN-based model,
yielding exact derivatives at low computational cost.
This formulation also naturally enables adaptive time-stepping strategies.
To illustrate this capability,
the imposed $\vec{H}$ field is designed to switch
from strongly saturated regions to low-field regions,
and an adaptive time-stepping strategy,
controlled by the predicted error $\varepsilon$,
is employed:
the time step $\Delta t$ is doubled when $\varepsilon < 25$~mT
and halved when $\varepsilon > 40$~mT,
with minimum and maximum time steps
of $1.25$~$\mu$s and $40$~$\mu$s, respectively.
When fully-saturated,
the anhysteretic contribution $\vec H_\text{anh}$ alone
governs the model response,
resulting in low predicted error $\varepsilon$
and allowing the adaptive scheme to increase the time step.
Conversely, at lower field levels,
the irreversible contribution $\tilde{\vec{H}}_\text{irr}$ becomes significant
and the predicted error $\varepsilon$ increases,
prompting the scheme to reduce the time step.
This adaptive strategy proves effective:
the predicted $\tilde{\vec{B}}$ is in close agreement
with the reference lamination solution $\vec{B}$.
Fig.~\ref{fig:testDtVar} also demonstrates the ability
of the model to extend out of its training range
in the saturation regime,
as $|\vec B|$ reaches 3.2~T.

%% file: FEM.tex
The magnetic vector potential $\vec{A}$ formulation is used for 2D FE simulations
of electrical machines because it reduces to a single scalar unknown
(the out of plane component of the vector potential)
discretized with nodal Lagrange elements (rather than edge elements),
makes $\div\vec B=0$ exact with flux lines as its contours,
lets prescribed winding currents enter trivially as a source while giving flux linkage and torque directly,
and handles the dominant non-conducting regions without any topological cuts \cite{Meunier2010}.

Let $\Omega$ denote the computational domain,
$\Omega_{\text{core}} \subset \Omega$ the laminated ferromagnetic core region,
$\Omega_s \subset \Omega$ the stranded conductor region
and $\vec j_s$ the imposed current density,
it seeks $\vec A$ such that

\begin{equation}
	\Big( \mu_0^{-1}\curl\vec A , \curl\vec A' \Big)
	_{\Omega \backslash \Omega_{\text{core}}}
	+\;\Big(
	\mu^{-1}\curl\vec A , \curl\vec A' \Big)
	_{\Omega_{\text{core}}}
	-\;\Big( \vec{j}_s , \vec A' \Big)_{\Omega_s}
	= 0
	\label{eq:nrformulation}
\end{equation}

\noindent
holds for all test function $\vec A'$ in the appropriate function space.
The coupling with the NN-based model is introduced by replacing the constitutive relation
$\mu^{-1}\curl\vec A = \mu^{-1}\vec B$ in $\Omega_{\text{core}}$
by the NN-based prediction $\tilde{\vec H}(\vec B)$.
Owing to the nonlinearity of $\tilde{\vec H}(\vec B)$,
the resulting problem is solved using a Newton-Raphson scheme.
At iteration $i$, the constitutive law is linearized around $\vec B_{i-1}$ as

\begin{equation}
	\tilde{\vec H}(\vec B_i)\approx\tilde{\vec H}(\vec B_{i-1})+\frac{\partial\tilde{\vec H}}{\partial\vec B}(\vec B_{i-1})\left(\vec B_i-\vec B_{i-1}\right),
\end{equation}

\noindent where $\tilde{\vec H}(\vec B_{i-1})$ is evaluated in a forward pass of the NN,
and the Jacobian $\partial\tilde{\vec H}/\partial\vec B(\vec B_{i-1})$
is computed by reverse-mode automatic differentiation.
Note that the GRU hidden state is kept fixed
during the Newton-Raphson iterations
and updated only upon convergence.
Consequently, once convergence has been reached,
the NN-based model must be queried one additional time to update its hidden state.
Alternatively, two hidden states can be stored per element in $\Omega_\text{core}$:
the input hidden state $S^{k-1}$,
which remains fixed during the Newton-Raphson iterations,
and the output hidden state $S^k_i$, returned by the GRU at each iteration,
which becomes the new input hidden state $S^k$ once convergence is reached.
The latter approach is adopted in this work,
requiring the storage of $2 \cdot 300 = 600$ double-precision numbers (4.8~kB)
per element in $\Omega_\text{core}$.

The proposed methodology is evaluated on two representative case studies,
both making use of the same NN-based model trained for stacks of M235-35A steel laminations.
Subsection \ref{sec:transfo} first considers a three-phase transformer,
and subsection \ref{sec:4e} then considers a synchronous reluctance motor.
In both applications, 
the models are discretized using the free and open-source software Gmsh~\cite{gmsh},
while the magnetic vector potential field $\vec A$
is discretized with first order elements
using the free and open-source FE solver GetDP \cite{getDP}.
All simulations reported in this work also perform element-wise integration
using a single Gauss quadrature point,
and consider relative and absolute residual tolerances
of $10^{-7}$ and $10^{-10}$, respectively.
The convergence behavior of the Newton-Raphson scheme
is discussed in in subsection \ref{sec:convergence},
while subsection \ref{sec:cost} analyses the computational cost.

\subsection{Three Phase Transformer}
\label{sec:transfo}

\begin{figure}[ht]
	\centering
	\includegraphics{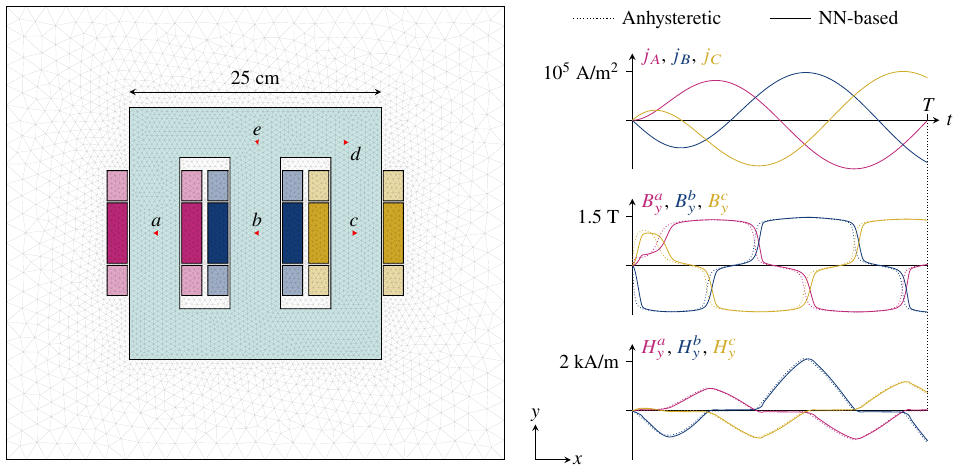}
	\caption{Left: Mesh of the considered three-phase transformer,
		with circa 5300 nodes and 10 500 elements, of which 4600 belong to $\Omega_\text{core}$.
		Right: Curves of the current densities $j_A$, $j_B$ and $j_C$ imposed in phase $A$, $B$ and $C$,
		and curves of the magnetic flux density $\vec B$ and magnetic field $\vec H$ along the $y$ direction in the center of each leg,
		resulting from a simulation with the proposed NN-based model (solid lines) or with an anhysteretic law (dotted lines).
		}
	\label{fig:transfo}
\end{figure}

For each phase $p \in \{A,B,C\}$ of the three-phase transformer,
the current density in the primary and secondary windings is imposed as
$j_p(t)=10^5\;(1-\exp(-2\pi ft))\;\sin(2\pi ft-\phi_p)$,
where $f=50$~Hz is the frequency,
$\phi_A=0^\circ$, $\phi_B=120^\circ$ and $\phi_C=240^\circ$ are phase lags,
and the factor $(1-\exp(-2\pi ft))$ introduces a gradual startup.
A homogeneous Dirichlet boundary condition is also imposed on the boundary of $\Omega$.
Fig.~\ref{fig:transfo} depicts the mesh of the analyzed three-phase transformer,
together with the curves of the imposed current density in each phase
and the corresponding $B_y(t)$ and $H_y(t)$ curves
at points $a$, $b$, and $c$ located at the center of each leg.
These curves result either from an anhysteretic simulation (dotted lines)
or from a simulation using the NN-based model (solid lines),
both running over 800 constant time steps for a single period.
The impact of hysteresis is visible on the $B_y$ curves which,
compared to the anhysteretic case,
are delayed due to remanent magnetization.

To evaluate the effect of time-step refinement on simulation accuracy,
two additional simulations are performed over a single period $T=1/f$
using the same trained NN-based model.
The first considers half less (400) constant time steps,
and the second uses an adaptive time step $\Delta t^k$.
At each time step $k$, the converged solution is accepted
if the maximum predicted error $\max_{\Omega}\varepsilon^k$ is lower than $100$~mT,
it is otherwise rejected and the time step is halved.
In case $\max_{\Omega}\varepsilon^k$ is lower than $70$~mT, the time step is doubled back,
with a maximum time step of $T/100=200~\mu$s,
and a minimum time step of $T/1600=12.5~\mu$s,
resulting in a simulation of 210 accepted time steps.
Note that this adaptive strategy is guided by the predicted error $\varepsilon^k$
rather than by an estimate of the local truncation error,
and should therefore be regarded as a heuristic approach.
Nevertheless, classical error-controlled time-stepping schemes remain applicable.

To measure the error introduced by the NN-based model,
the $\vec H$ sequences (one per element in $\Omega_\text{core}$)
obtained during the simulation with 800 time steps
are injected into the lamination model.
The resulting sequences, noted $\bar{\vec B}$,
serve as target and can be compared to the $\vec B$ sequences
obtained directly during the three simulations using the NN-based model.
They are used to compute the mismatch $|\vec B - \bar{\vec B}|$,
which is a direct measure of the simulation accuracy.
If $|\vec B - \bar{\vec B}|$ was zero at each Gauss point and at each time step,
it would mean the NN-based model is a perfect substitute to the lamination model.
The same results could therefore be obtained using the lamination model as material law,
using sufficiently fine temporal refinement
and at a computational cost and complexity largely superior.
\begin{figure}[!ht]
	\centering
	\includegraphics[]{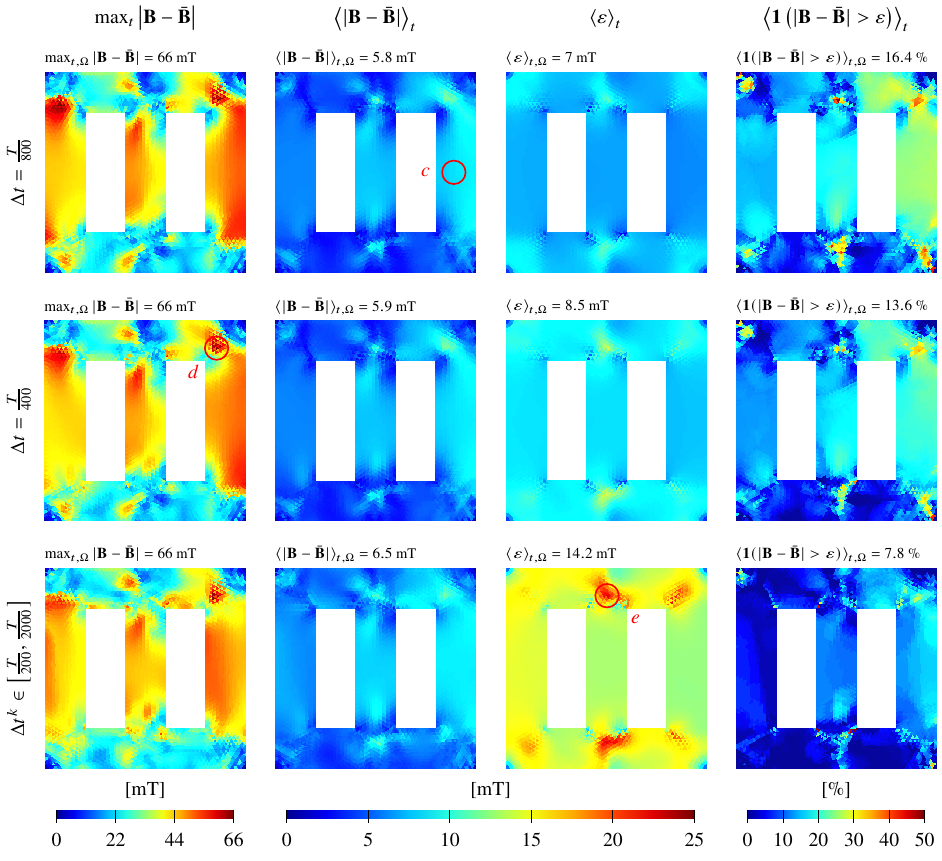}
	\caption{Spatial distributions of the
		maximum-in-time mismatch (first column),
		time-averaged mismatch (second column),
		time-averaged predicted error (third column),
		and exceedance (last column),
		for the three NN-based simulations.
	}
	\label{fig:mapsTransfo}
\end{figure}

The first two columns of Fig.~\ref{fig:mapsTransfo} are used to assess the simulation accuracy.
The first column displays the spatial distribution of the maximum-in-time mismatch $\max_t |\vec{B} - \bar{\vec{B}}|$,
while the second displays the spatial distribution of the time-averaged mismatch $\langle |\vec{B} - \bar{\vec{B}}| \rangle_t$,
both computed element-wise over the ferromagnetic core region $\Omega_{\text{core}}$,
and for the three aforementioned NN-based simulations.
For all three simulations,
the maximum-in-time mismatch remains below 66~mT
throughout $\Omega_{\text{core}}$,
while the mean mismatch $\langle |\vec{B} - \bar{\vec{B}}| \rangle_{t,\Omega}$,
averaged over time and space,
is about 6~mT.
When compared locally with the norm of $\vec{B}$,
these mismatches correspond to relative values
$\langle |\vec{B} - \bar{\vec{B}}| \rangle_t / \langle |\vec{B}| \rangle_t$
that remain below 5.4~\% throughout $\Omega_{\text{core}}$,
with a spatial average
$\langle \langle |\vec{B} - \bar{\vec{B}}| \rangle_t / \langle |\vec{B}| \rangle_t \rangle_{\Omega}$
of about 0.7~\%.

Without access to the lamination model,
the mismatch $|\vec B - \bar{\vec B}|$ cannot be computed,
and one should rely exclusively on the predicted error $\varepsilon$.
The third column of Fig.~\ref{fig:mapsTransfo} shows the spatial distribution of its time average $\langle\varepsilon\rangle_t$.
It is observed that, except in the vicinity of point $c$,
$\langle\varepsilon\rangle_t$ remains larger than the time-averaged mismatch $\langle|\vec B-\bar{\vec B}|\rangle_t$.
Furthermore, locations where the time-averaged mismatch is larger
correspond to locations where the time-averaged predicted error is larger,
showing the ability of the NN to detect challenging regions.

The time-averaged mismatch $\langle|\vec B-\bar{\vec B}|\rangle_t$
and time-averaged predicted error $\langle\varepsilon\rangle_t$ however
provide only aggregate measures over time and do not indicate
whether the predicted error consistently bounds the actual mismatch,
or if the mismatch frequently exceeds the predicted error.
The last column of Fig.~\ref{fig:mapsTransfo} reports the percentage of time
the mismatch is larger than the predicted error,
hereafter referred as the exceedance $\langle\vec 1(|\vec{B}-\bar{\vec{B}}|>\varepsilon)\rangle_t$.
The mean exceedance $\langle\vec 1(|\vec{B}-\bar{\vec{B}}|>\varepsilon)\rangle_{t,\Omega}$ is obtained by further averaging over space.
It reaches 16.4~\%, 13.6~\% and 7.8~\%
for respectively the 800-time-step, 400-time-step and variable-time-step simulations,
showing the predicted error indeed remains larger than the mismatch most of the time.
The mean exceedance decreases to 4.9~\%, 4~\% and 2.2~\%
when considering twice the predicted error,
and to 2~\%, 1.9~\% and 0.9~\% when considering three times the predicted error.

\begin{figure}[!htbp]
	\centering
	\includegraphics[]{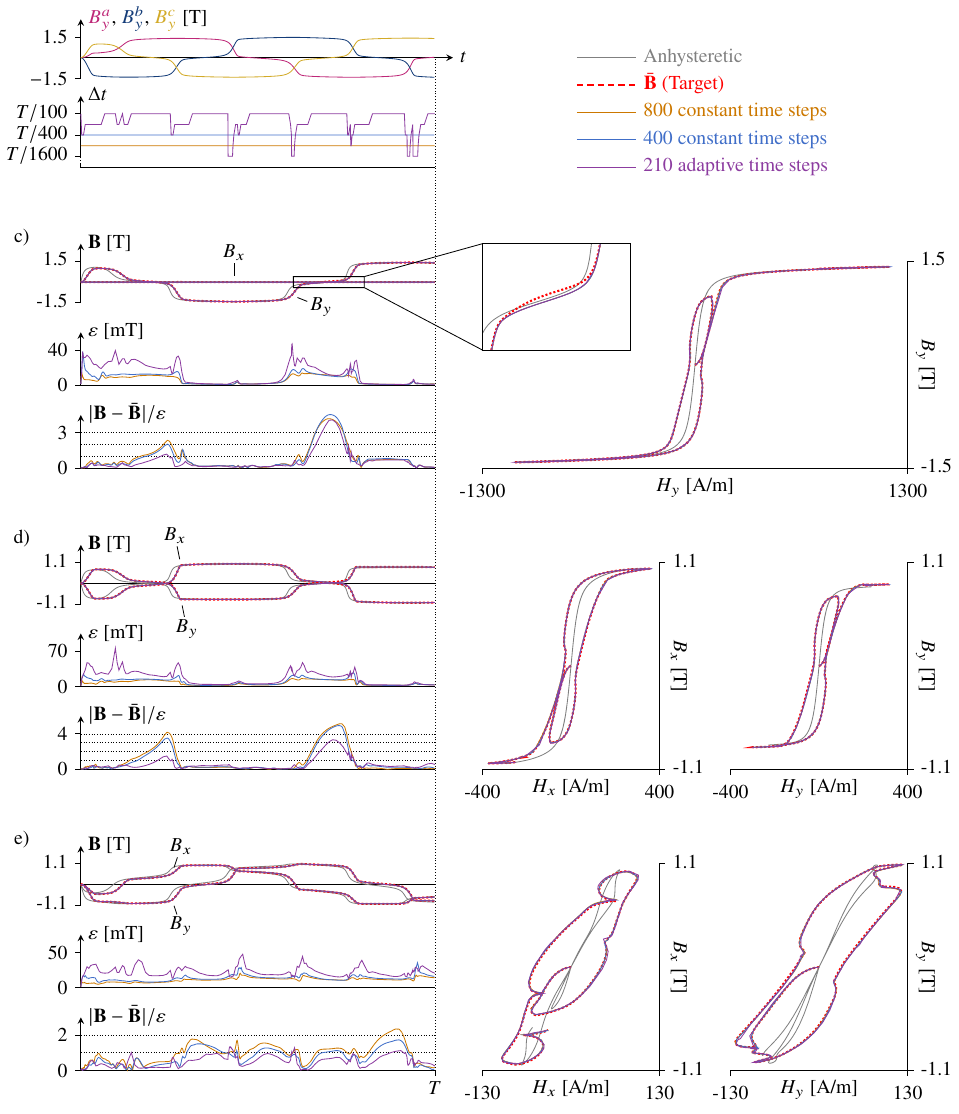}
	\caption{Comparison of the $\vec B$,
		$\varepsilon$,
		$|\vec B-\bar{\vec B}|/\varepsilon$,
		and $\vec{B}(\vec{H})$ curves
		obtained during the NN-based simulations with
		800 constant time steps,
		400 constant time steps,
		and 210 adaptive time steps,
		at the points 
		$c$ (time-averaged mismatch greater than time-averaged predicted error),
		$d$ (largest maximum-in-time mismatch),
		and $e$ (largest time-averaged predicted error)
		identified in Fig.~\ref{fig:transfo} and \ref{fig:mapsTransfo}.
		The target $\bar{\vec B}$ curves are given by the lamination model,
		evaluated on the $\vec H$ curves obtained during the 800-time-step NN-based simulation.
		At the top of the figure,
		the $\Delta t$ curves are also depicted, together with the $B_y$ curves obtained
		at the center of each leg of the three-phase transformer (points $a$, $b$, and $c$).
	}
	\label{fig:curvesTransfo}
\end{figure}

While the previous analysis focuses on global and time-averaged quantities,
it is also instructive to examine the behavior at specific points.
Fig.~\ref{fig:curvesTransfo} shows the time evolution of
$\vec B$, $\varepsilon$, and $|\vec B-\bar{\vec B}|/\varepsilon$,
together with the corresponding $\vec B(\vec H)$ curves,
at points $c$, $d$, and $e$, identified in Fig.~\ref{fig:transfo} and~\ref{fig:mapsTransfo}.
Point $c$ stands in the middle of the region where
the time-averaged mismatch is greater than the time-averaged predicted error,
while point $d$ exhibits the worst maximum-in-time mismatch
and point $e$ showcases the largest time-averaged predicted error.
Primarily focusing on the $\vec B$ and $\vec B(\vec H)$ curves,
it is observed that
the NN curves closely match the target curves.
The $B_y$ curve of point $c$ also shows that the higher time-average mismatch
does not originate from a discrepancy in shape,
but rather results from a slight amplitude offset between the target and prediction curves.
This curve explains why the time-averaged mismatch is slightly larger
than the time-averaged predicted error in the vicinity of point $c$.
Despite this deviation, the predicted $B_y(H_y)$ loop still
accurately reproduces the reference curve.
By comparing the $\varepsilon$ and $\vec B$ curves,
it appears that $\varepsilon$ is minimum when $|\vec B|$ is sufficiently large to reach saturation,
in which case the model prediction mainly relies on the anhysteretic contribution $\vec H_\text{anh}$.
Fig.~\ref{fig:curvesTransfo} also depicts the evolution of the time step for the adaptive-time-step simulation,
and the $B_y$ curves at the center of each of the transformer leg.
The time step is decreased whenever a transformer leg
switches from nearly-zero field to saturation or inversely,
reflecting a rise in the predicted error during these transitions.
Eventually, the $|\vec B-\bar{\vec B}|/\varepsilon$ curves reveal that the previously reported exceedance
translates locally into ratios that stay close to unity for most of the time horizon.

\subsection{Synchronous Reluctance Motor}
\label{sec:4e}

\begin{figure}[!htbp]
	\centering
	\includegraphics[scale=1.0]{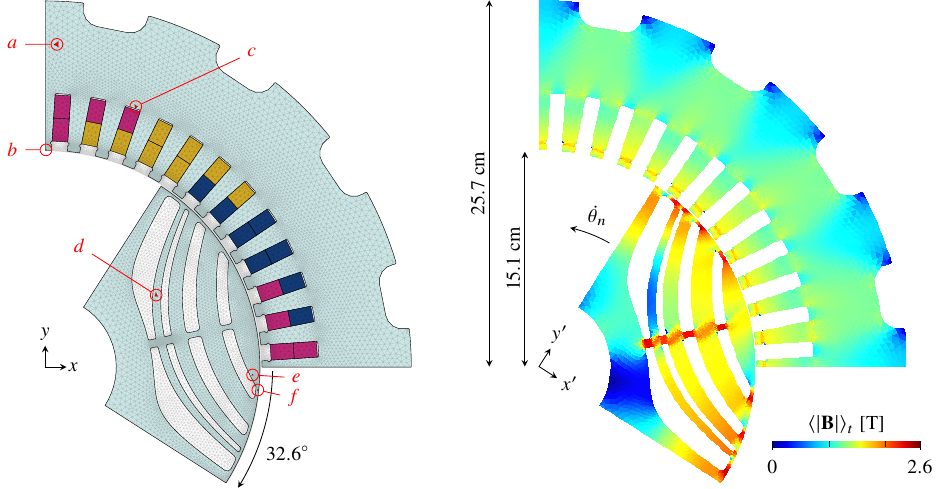}
	\caption{Left: Mesh of the considered synchronous reluctance motor,
		with circa 14700 nodes and 29600 elements, of which 17000 belong to $\Omega_\text{core}$.
		Right: Norm of $\vec B$, averaged over time.}
	\label{fig:4e}
\end{figure}

The considered synchronous reluctance motor is supplied with three-phase sinusoidal currents of $217.31~A_{\mathrm{RMS}}$,
delivering a power of $220$~kW.
It has two pole pairs and operates at a frequency of $65$~Hz,
resulting in a nominal speed of $\dot{\,\theta_n}=1950$~rpm.
The flux barriers in the rotor result from a shape optimization process
aimed at minimizing torque ripple~\cite{DeGreef2021}.
Fig.~\ref{fig:4e} presents the considered mesh ---
which relies on a moving band technique
to handle the relative motion between stator and rotor ---
together with a map of the time-averaged magnetic-flux-density amplitude $\langle|\vec B|\rangle_t$.
It also introduces two coordinate systems:
a fixed $xy$ frame and a rotating $x'y'$ frame attached to the rotor.
When evaluating the NN-based model,
the rotor sequences must be expressed according to the $x'y'$ frame,
in order to eliminate artificial time variations due to rotation.

As with the transformer,
the target $\bar{\vec B}$ sequences (one per element in $\Omega_{\text{core}}$)
are obtained by first running an NN-based simulation,
and subsequently feeding the resulting $\vec H$ sequences to the lamination model.
This simulation uses 4800 constant time steps to model a single period,
a deliberately large number to ensure a sufficiently fine temporal resolution
and accurately capture rapid variations due to rotor--stator tooth alignment.
Simulations with a coarser temporal resolution may lack precision 
in the predicted $\vec H$ sequences and,
although the impact on subsequent NN predictions remains limited,
it could drive the EB hysteresis model cells
into a spurious state from which they cannot automatically recover,
ultimately compromising the accuracy of the target $\bar{\vec B}$ sequences.

At first, these $\bar{\vec B}$ sequences are compared against the $\vec B$ sequences
obtained from an NN-based simulation with 1200 time steps for a single period.
Unlike the simulations of the transformer,
the three-phase currents are imposed directly as sinusoidal waveforms at their nominal amplitude,
without any ramp-up phase.
As a result, the first time step is more challenging to converge,
since the solution must evolve from an initially 
completely demagnetized state along the initial magnetization curve.
To facilitate convergence, a relaxation factor of 0.8 is applied for this initial time step,
and no strict limit is imposed on its number of iterations.

\begin{figure}[!htb]
	\centering
	\includegraphics[scale=1.0]{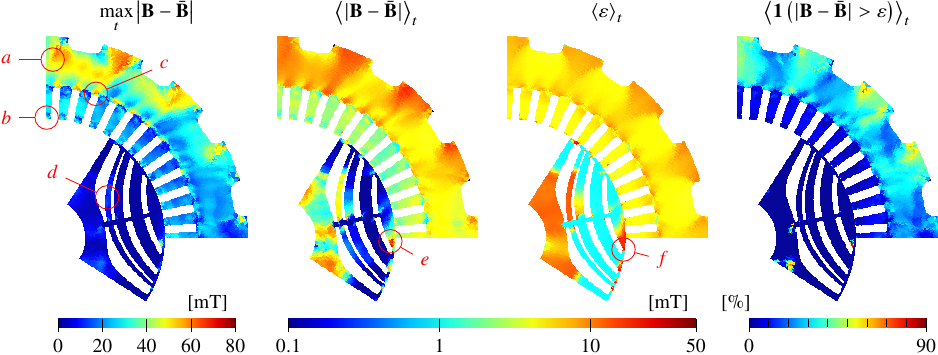}
	\caption{Spatial distributions of the
		maximum-in-time mismatch (first column),
		time-averaged mismatch (second column),
		time-averaged predicted error (third column),
		and exceedance (last column),
		for the 1200-time-step NN-based simulation.}
	\label{fig:maps4e}
\end{figure}

Fig.~\ref{fig:maps4e} presents the spatial distributions of the
maximum-in-time mismatch $\max_t|\vec B-\bar{\vec B}|$,
time-averaged mismatch $\langle|\vec B-\bar{\vec B}|\rangle_t$,
time-averaged predicted error $\langle\varepsilon\rangle_t$,
and exceedance $\langle\vec 1(|\vec B-\bar{\vec B}|>\varepsilon)\rangle_t$.
The peak mismatch $\max_{t,\Omega} |\vec{B} - \bar{\vec{B}}|$ reaches 73~mT at point $c$,
while the time-averaged mismatch remains below 27.5~mT
throughout $\Omega_{\text{core}}$, reaching its maximum at point $e$.
The time-averaged predicted error attains its maximum
at the rotor air-gap interface (point $f$),
where the field undergoes the most abrupt variations,
ranging from strong saturation when aligned with a stator tooth,
to nearly zero otherwise.
The predicted error also remains strictly above 1~mT,
as discussed in Section~\ref{sec:NNLL},
and, except for a region near point $a$,
consistently bounds the time-averaged mismatch.
The mean relative mismatch is here computed as
$\langle \langle |\vec{B} - \bar{\vec{B}}| \rangle_t / (\langle |\vec{B}| \rangle_t + \epsilon) \rangle_{\Omega}$,
where $\epsilon = 1$~mT is introduced to avoid near-zero divisions
in regions where $\langle|\vec{B}|\rangle_t$ vanishes (see Fig.~\ref{fig:4e}).
The resulting mean relative mismatch is 0.5~\%.

\begin{figure}[!htbp]
	\centering
	\includegraphics{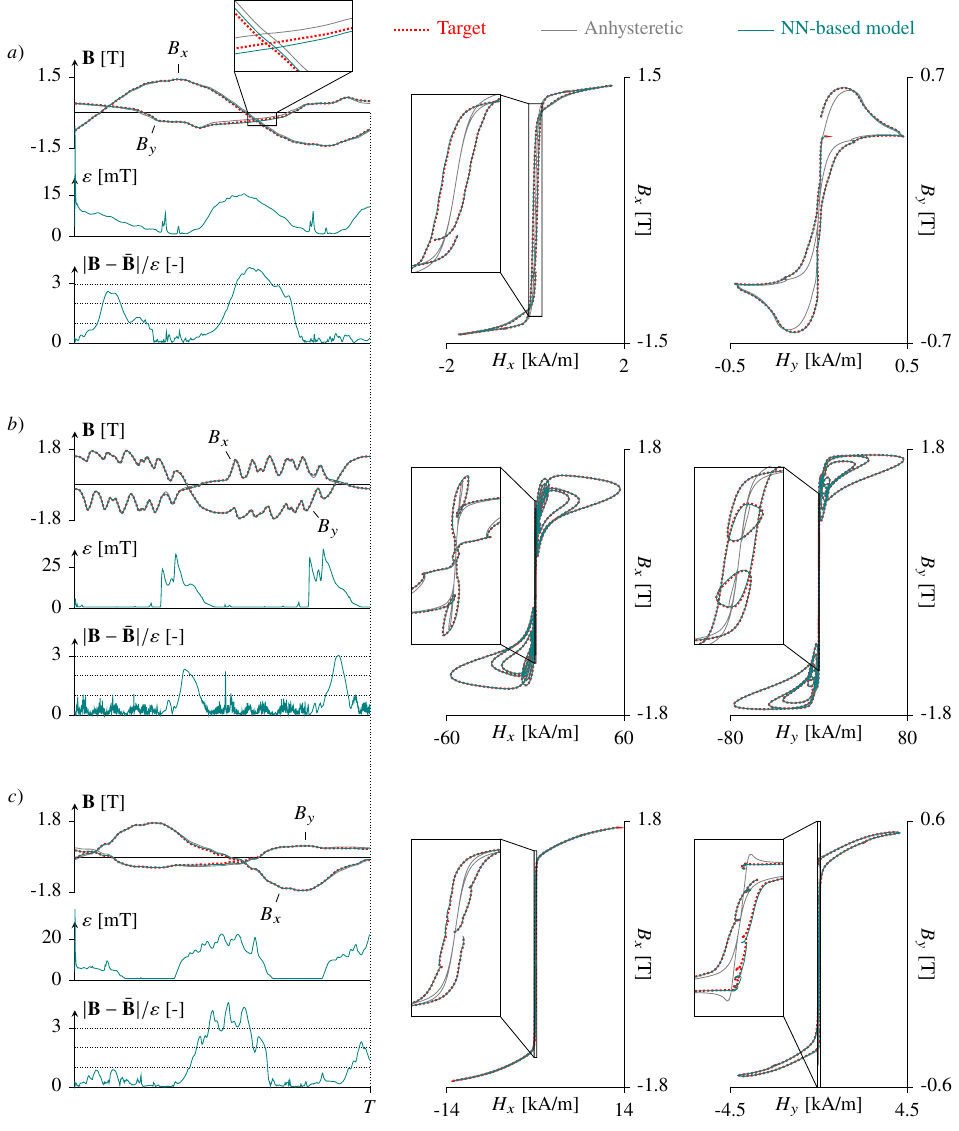}
	\caption{Comparison of the $\vec B$,
		$\varepsilon$,
		$|\vec B-\bar{\vec B}|/\varepsilon$,
		and $\vec{B}(\vec{H})$ curves 
		obtained during the NN-based simulation with 1200 time steps,
		at the stator points $a$, $b$ and $c$,
		identified in Fig.~\ref{fig:4e} and \ref{fig:maps4e}.
		The target $\bar{\vec B}$ curves are given by the lamination model,
		evaluated on the $\vec H$ curves obtained during a 4800-time-step NN-based simulation.}
	\label{fig:curvesStator}
\end{figure}

\begin{figure}[!htbp]
	\centering
	\includegraphics[]{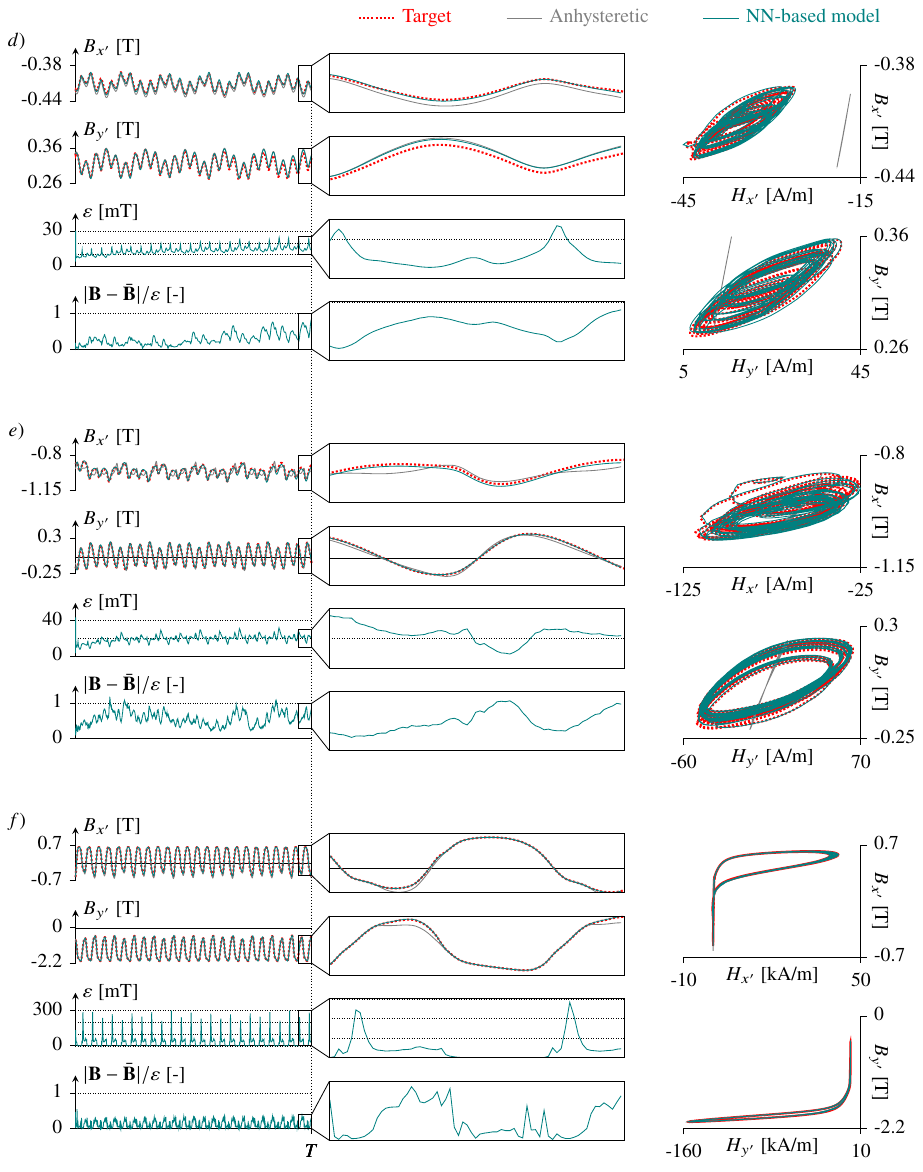}
	\caption{Comparison of the $\vec B$,
		$\varepsilon$,
		$|\vec B-\bar{\vec B}|/\varepsilon$,
		and $\vec{B}(\vec{H})$ curves 
		obtained during the NN-based simulation with 1200 time steps,
		at the rotor points $d$, $e$ and $f$,
		identified in Fig.~\ref{fig:4e} and \ref{fig:maps4e}.
		The target $\bar{\vec B}$ curves are given by the lamination model,
		evaluated on the $\vec H$ curves obtained during a 4800-time-step NN-based simulation.}
	\label{fig:curvesRotor}
\end{figure}

In Fig.~\ref{fig:curvesStator} and \ref{fig:curvesRotor},
the $\vec B$, $\varepsilon$, $|\vec B-\bar{\vec B}|/\varepsilon$, and $\vec B(\vec H)$
curves are displayed at points $a$, $b$ and $c$ in the stator,
and at points $d$, $e$ and $f$ in the rotor.
Similarly to point $c$ in the transformer,
the $\vec B$ and $\bar{\vec B}$ curves at point $a$ exhibit a slight amplitude shift,
which explains the localized larger error near point $a$.
For all the other curves, $\vec B$ and $\bar{\vec B}$ are in close agreement.

To assess the sensitivity of the method to reduced temporal resolution,
three additional NN-based simulations are performed with 600, 300, and 150
nominal time steps per period.
Note that, in practice, an adaptive time-stepping strategy is employed:
except for the first time step, the time step is halved whenever convergence is not achieved within 20 iterations,
and doubled when convergence is reached in fewer than 10 iterations.
Table \ref{tab:4e} reports the
peak mismatch $\max_{t,\Omega} |\vec{B} - \bar{\vec{B}}|$,
mean mismatch $\langle |\vec{B} - \bar{\vec{B}}| \rangle_{t,\Omega}$,
and mean predicted error $\langle\varepsilon\rangle_{t,\Omega}$
at the stator and at the rotor
for the aforementioned NN-based simulations.
It is first worth noting that the simulations maintain good accuracy on average,
as the mean stator and rotor mismatches increase only modestly,
from 6.1 to 6.9~mT and 1.8 to 2.6~mT, respectively,
when reducing the number of nominal time steps from 1200 to 150.
Moreover, the mean and peak mismatches remain nearly unchanged
between 1200 and 300 nominal time steps,
suggesting that a resolution of 300 time steps is already sufficient.
This observation is further supported by the behavior of the mean predicted error,
which increases only slightly from 1200 to 300 time steps,
but rises more markedly from 300 to 150 time steps.

\begin{table}[!htbp]
	\caption{Peak mismatch, mean mismatch, and mean predicted error
		at the stator and rotor for the NN-based simulations with 1200, 600, 300, and 150 nominal time steps per period.
	}
	\label{tab:4e}
	\centering
	\begin{tabular}{c|ccc|ccc}
		\noalign{\vskip 3pt}
		\noalign{\hrule height 1.2pt}
		\noalign{\vskip 3pt}
		Nominal number
		& \multicolumn{3}{c|}{Stator [mT]} & \multicolumn{3}{c}{Rotor [mT]}\\
		of time steps &$\max_{t,\Omega} \lvert \vec{B}-\bar{\vec{B}} \rvert$&$ \langle \lvert \vec{B}-\bar{\vec{B}} \rvert \rangle_{t,\Omega}$&$ \langle \varepsilon \rangle_{t,\Omega}$ &$ \max_{t,\Omega} \lvert \vec{B}-\bar{\vec{B}} \rvert$&$ \langle \lvert \vec{B}-\bar{\vec{B}} \rvert \rangle_{t,\Omega}$&$ \langle \varepsilon \rangle_{t,\Omega}$\\
		\noalign{\vskip 3pt}
		\noalign{\hrule height 1.2pt}
		\noalign{\vskip 3pt}
		1200 & 73.0 & 6.1 & 5.8 & 47.3 & 1.8 & 5.5\\
		600 & 77.5 & 6.1 & 7.0 & 52.5 & 1.9 & 7.0\\
		300 & 82.6 & 6.2 & 11.3 & 204.9 & 2.1 & 11.2\\
		150 & 320.2 & 6.9 & 26.0 & 701.3 & 2.6 & 20.7\\
		\noalign{\vskip 3pt}
		\noalign{\hrule height 1.2pt}
	\end{tabular}
\end{table}

\begin{figure}[!htb]
	\centering
	\includegraphics[]{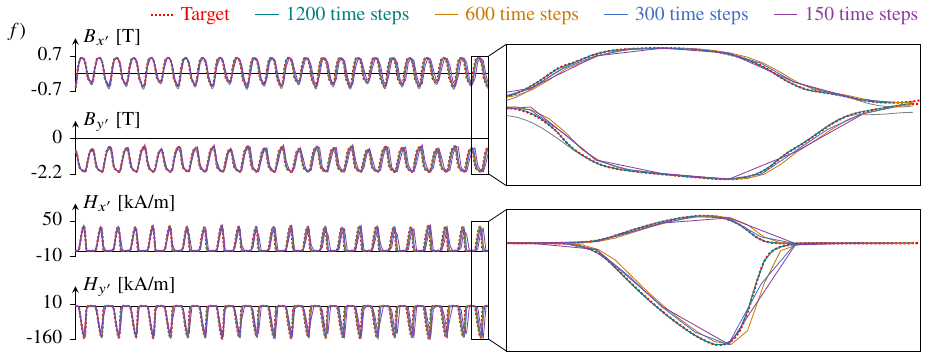}
	\caption{Comparison of the $\vec{B}$ and $\vec H$ curves 
		obtained during the NN-based simulations with 1200, 600, 300 and 150 nominal time steps,
		at the rotor point $f$.}
	\label{fig:curveRotorAirgap}
\end{figure}

\begin{figure}[!htb]
	\centering
	\includegraphics[]{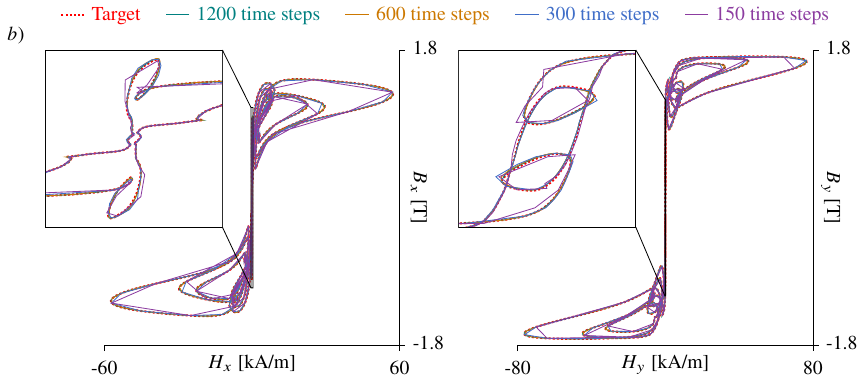}
	\caption{Comparison of the $\vec{B}(\vec{H})$ curves 
		obtained during the NN-based simulations with 1200, 600, 300 and 150 nominal time steps,
		at the stator point $b$.}
	\label{fig:curves4enT}
\end{figure}

As the temporal resolution is reduced,
the peak error increases more markedly at the rotor than at the stator,
which can be explained by the coarser temporal discretization of the rotor sequences.
Indeed, with 48 stator teeth and two pole pairs,
the $24^\text{th}$ harmonic dominates at the rotor
and is sampled by only about 6 time steps
in the simulation with 150 nominal time steps per period.
Nevertheless, Fig.~\ref{fig:curveRotorAirgap},
which shows the $\vec B$ and $\vec H$ fields
at point $f$ in the rotor for the 1200, 600, 300, and 150 nominal-time-step simulations,
demonstrates the ability of the NN-based model to handle such poorly resolved temporal variations.
This robustness is particularly valuable during the design phase of electrical machines,
where coarse time discretizations are often preferred to reduce computational costs.
In this regime, the model strongly benefits from the anhysteretic contribution $\vec H_\text{anh}$,
which provides a physically consistent base to the NN predictions,
regardless of the temporal discretization.
Fig.~\ref{fig:curves4enT} further compares these NN-based simulations
by depicting the $\vec B(\vec H)$ curves at point $b$ in the stator.
Owing to the contribution of $\vec H_\text{anh}$,
the NN-based model provides perfect predictions in the strongly saturated regime,
even for coarse temporal discretizations.
At lower field levels, although some noise is observed,
even the 150-nominal-time-step simulation remains capable of capturing the minor hysteresis loops.


\newpage

\subsection{Convergence of the Newton-Raphson scheme}
\label{sec:convergence}

\begin{figure}[ht]
	\centering
	\includegraphics[]{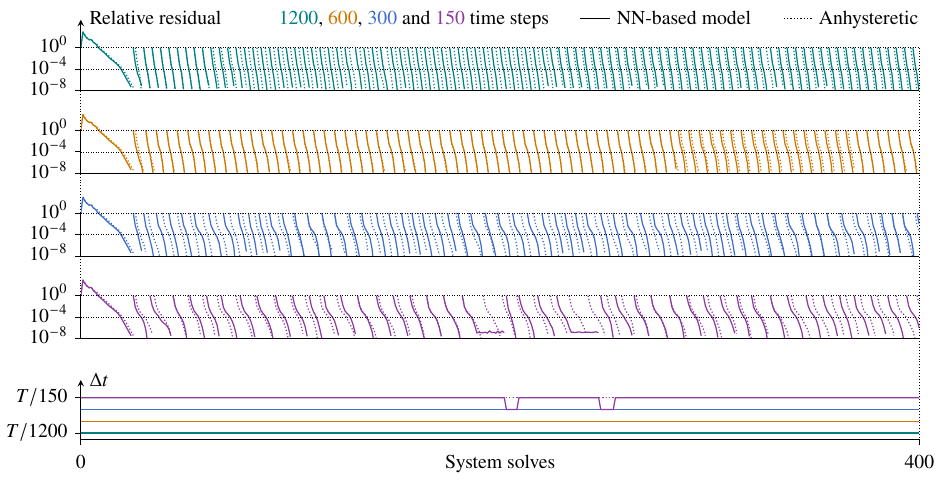}
	\caption{Convergence history of the nonlinear Newton-Raphson iterations,
		with the anhysteretic material law and with the NN-based model,
		considering the 1200, 600, 300 and 150 nominal-time-step simulations
		of the synchronous reluctance machine.}
	\label{fig:res4e}
\end{figure}

Fig.~\ref{fig:res4e} presents the convergence behavior
of the Newton-Raphson scheme for the synchronous reluctance motor simulations
performed with 1200, 600, 300, and 150 nominal time steps,
using either the NN-based surrogate model or the anhysteretic law.
Except for a limited number of instances in the 150-time-step case,
where the NN-based simulation reduces its time step
after reaching the maximum of 20 Newton iterations,
both models exhibit nearly identical convergence behavior.
This close superposition indicates that the NN-based model preserves
the numerical behavior of the anhysteretic model.
The initial time step also requires more Newton iterations to converge since,
as previously mentioned,
the three-phase currents are imposed directly as sinusoidal waveforms at their nominal amplitude.
Consequently, the solution must evolve from a completely demagnetized state
and stabilize along the first magnetization curve.

\newpage
\subsection{Computational Cost}
\label{sec:cost}

All simulations were carried out on a MacBook Pro laptop equipped with an Apple M1 Max chip,
using GetDP for the FE computations \cite{getDP}
and Python to evaluate either the anhysteretic law or the NN-based surrogate model.
In practice, the transformer simulations run in 12.7, 6.8, and 4.3 minutes
for the 800-, 400-, and adaptive-time-step cases respectively,
while the synchronous reluctance motor simulations run in
55.4, 34.2, 20.9 and 14.8 minutes
for 1200, 600, 300, and 150 nominal time steps respectively.
Tables~\ref{tab:timingsTransfo} and \ref{tab:timings4e} provide
a detailed breakdown of these timings for the transformer and synchronous reluctance motor simulations,
decomposed into three contributions:
evaluation of the material law in Python,
assembly and solution of the system in GetDP,
and input/output (I/O) operations corresponding to data exchange between Python and GetDP.
This exchange is currently implemented via text files written to and read from disk,
which, while functional, accounts for approximately 20~\% of the total simulation time.
In future work, this file-based communication will be replaced
by shared-memory communication to remove this overhead.

\begin{table}[!ht]
	\caption{Comparison of computation times and number of system solves
		for the transformer simulations.
		``Total time steps'' denotes all attempted time steps
		(accepted and rejected),
		whereas ``effective time steps'' counts only the accepted ones;
		the two coincide for fixed time-step simulations.
	}
	\label{tab:timingsTransfo}
	\centering
	\begin{tabular}{ccc|cc|cccc|cc}
		\noalign{\vskip 3pt}
		\noalign{\hrule height 1.2pt}
		\noalign{\vskip 3pt}
		& \multicolumn{2}{c|}{Time steps} & \multicolumn{2}{c|}{System solves} & \multicolumn{4}{c|}{Time [s]} & \multicolumn{2}{c}{Time/system solve [s]} \\
		& Total & Effective & Total & /Step & Python & GetDP & I/O & Total & Python & GetDP\\
		\noalign{\vskip 3pt}
		\noalign{\hrule height 1.2pt}
		\noalign{\vskip 3pt}
		\multirow{ 4}{*}{\rotatebox{90}{$\;\;\;$NN-b.}}
		& 800 & 800 & 3631 & 4.54 & 292 & 311 & 158 & 762 & 0.08 & 0.09\\
		& 400 & 400 & 1944 & 4.86 & 157 & 167 & 83 & 409 & 0.08 & 0.09\\
		& 240 & 210 & 1243 & 5.18 & 93 & 109 & 53 & 256 & 0.08 & 0.09\\
		\noalign{\vskip 3pt}
		\noalign{\hrule height 1.2pt}
		\noalign{\vskip 3pt}
		\multirow{ 4}{*}{\rotatebox{90}{$\quad\;\;\;\;$Anh.}}
		& 800 & 800 & 3321 & 4.15 & 2 & 282 & 138 & 424 & / & 0.09\\
		& 400 & 400 & 1751 & 4.38 & 1 & 147 & 72 & 221 & / & 0.08\\
		\noalign{\vskip 3pt}
		\noalign{\hrule height 1.2pt}
	\end{tabular}
\end{table}

\begin{table}[!ht]
	\caption{Comparison of computation times and number of system solves
		for the synchronous-reluctance-motor simulations.
		``Nominal time steps'' refers to the target number of time steps per period,
		whereas ``effective time steps'' denotes the actual number of accepted time steps
		after adaptive refinement;
		the two differ only when additional intermediate time steps are introduced
		following a failed convergence attempt.
	}
	\label{tab:timings4e}
	\centering
	\begin{tabular}{ccc|cc|cccc|cc}
		\noalign{\vskip 3pt}
		\noalign{\hrule height 1.2pt}
		\noalign{\vskip 3pt}
		& \multicolumn{2}{c|}{Time steps} & \multicolumn{2}{c|}{System solves} & \multicolumn{4}{c|}{Time [s]} & \multicolumn{2}{c}{Time/system solve [s]}\\
		& Nominal & Effective & Total & /Step & Python & GetDP & I/O & Total & Python& GetDP\\
		\noalign{\vskip 3pt}
		\noalign{\hrule height 1.2pt}
		\noalign{\vskip 3pt}
		\multirow{ 4}{*}{\rotatebox[origin=c]{90}{NN-based}}
		& 1200 & 1200 & 4860 & 4.05 & 1376 & 1180 & 766 & 3323 & 0.28 & 0.24\\
		& 600 & 600 & 3007 & 5.01 & 844 & 738 & 467 & 2050 & 0.28 & 0.25\\
		& 300 & 300 & 1829 & 6.10 & 517 & 452 & 282 & 1252 & 0.28 & 0.25\\
		& 150 & 153 & 1290 & 8.43 & 365 & 323 & 197 & 886 & 0.28 & 0.25\\
		\noalign{\vskip 3pt}
		\noalign{\hrule height 1.2pt}
		\noalign{\vskip 3pt}
		\multirow{ 4}{*}{\rotatebox{90}{Anhyst.$\;$}}
		& 1200 & 1200 & 4912 & 4.09 & 6 & 1171 & 739 & 1917 & / & 0.24 \\
		& 600 & 600 & 3001 & 5.00 & 3 & 726 & 452 & 1182 & / & 0.24 \\
		& 300 & 300 & 1849 & 6.16 & 2 & 450 & 277 & 730 & / & 0.24\\
		& 150 & 150 & 1259 & 8.39 & 1 & 310 & 189 & 501 & / & 0.25\\
		\noalign{\vskip 3pt}
		\noalign{\hrule height 1.2pt}
	\end{tabular}
\end{table}

For both applications,
the time required to evaluate the NN-based model in Python
is approximately that needed to assemble and solve the system in GetDP.
Compared to the anhysteretic simulations,
for which the time spent in Python is negligible,
and assuming that the I/O overhead would be removed through improved communication,
a simulation using the NN-based model is therefore limited to roughly twice
the cost of an anhysteretic simulation.
Beyond these absolute timings, the computational cost scales linearly with problem size.
The number of elements in $\Omega$ increases by a factor of 2.8
from the transformer to the synchronous reluctance motor (10500 to 29600),
and the GetDP assembly and solve time per system solve
grows by a consistent factor of 2.8 (0.09 to 0.25~s).
The same holds for the Python evaluation,
where a 3.7-fold increase in $\Omega_\text{core}$ elements (4600 to 17000)
yields a 3.7-fold increase in evaluation time (0.08 to 0.28~s).

%% file: conclusion.tex
This paper has discussed the 2D FE modeling
of electrical machines with laminated ferromagnetic cores.
The results show that
homogenization of the laminated ferromagnetic core behavior
via a surrogate NN-based model offers
key implementation and numerical advantages 
over direct computational homogenization:
i) The automatic differentiation capabilities of NNs
allow an exact evaluation of the NN-based ferromagnetic model Jacobian,
eliminating the need for finite differences.
Furthermore, since the proposed NN is smooth by construction,
owing to its differentiable activation functions,
the resulting Jacobian matrices are smooth,
yielding Newton-Raphson convergence comparable to that obtained
with conventional anhysteretic constitutive laws.
In contrast, Jacobians obtained using direct homogenization
would inherit the numerical noise of finite differences
and could be affected by the non-smoothness of the EB hysteresis model.
ii) The NN allows coupling an $\vec H$-conform mesoscopic-scale model
(to use the fast vector-play EB approximation)
with an efficient $\vec B$-conform macroscopic machine model.
iii) The NN-based model inherits the accuracy of the fine temporal discretization
used during dataset generation
and largely preserves this accuracy at inference,
even when coarser time steps are used at the macroscopic scale.
Furthermore, after training on variable time steps,
the NN-based model seamlessly handles variable time increments,
so that macroscopic-scale adaptive time-stepping strategies are naturally supported.
Achieving the same with a direct homogenization approach
would require maintaining a fine temporal discretization at the mesoscopic scale
throughout the simulation,
thereby incurring the full computational cost of fine mesoscopic-scale solves \cite{Niyonzima2016Waveform}
iv) From the perspective of the FE solver,
the proposed constitutive model behaves as a single function call.
The magnetic history required to describe hysteresis
is entirely encoded in the NN hidden state,
eliminating the need to explicitly manage
the internal variables of an underlying mesoscopic-scale model.
v) Once trained for a given lamination (material grade and sheet thickness),
the proposed surrogate enables computationally efficient FE simulations.
Although extending the approach to a new lamination
requires generating a new dataset
(approximately 4 days on a single CPU core, but embarrassingly parallel)
and retraining the network (approximately 2 days),
the model can subsequently be used in a wide range of FE simulations,
with computational times roughly twice those of conventional anhysteretic simulations.
Moreover, the predicted error returned by the NN-based model
provides an immediate accuracy measure of the FE simulations.

Several directions are left for future work.
The present study considers sinusoidal current excitations,
whereas motors are often operated using pulse-width modulation (PWM);
assessing the performance of the surrogate model under such conditions is a natural next step.
The NN architecture could also be extended to predict ferromagnetic core losses
and to account for temperature effects,
following an approach similar to~\cite{denis2026acceleration},
where NN-based models are used
for AC loss prediction in superconducting conductors.
On a broader scale, the method currently assumes isotropic magnetic properties;
extending it to anisotropic laminations, such as grain-oriented electrical steels,
would require adapting both the dataset generation and the network architecture.
A 3D extension would also be needed to accurately capture eddy-current paths
near edges and geometric discontinuities.

%% file: resampling.tex
\makeatletter
\NewDocumentCommand{\LeftComment}{s m}{%
	\Statex \IfBooleanF{#1}{\hspace*{\ALG@thistlm}}\(\triangleright\) #2}
\makeatother

The resampling algorithm is designed to produce sequences
with irregular time steps of varying density.
It first thins the source sub-sequence by retaining only a random subset of points,
then re-enriches it by inserting linearly interpolated points,
whose distribution can range from spatially uniform to highly localized.

\vspace{0.3cm}
\hrule
\vspace{0.1cm}
\begin{algorithmic}
	\Statex Inputs:
	\Statex \hspace{1em} $\vec B_\text{src}$: source sub-sequence
	\Statex \hspace{1em} $n_\text{src}$: length of the source sub-sequence
	\Statex \hspace{1em} $n_\text{ins}$: number of points to insert
	\Statex \hspace{1em} $\bar{g}$: maximum gap between two consecutive kept source points
	\Statex \hspace{1em} $\bar{m}$: maximum number of points inserted per segment
	\Statex \hspace{1em} $\delta_\text{min}$: minimum spacing between consecutive output points
	
	\Statex Output:
	\Statex \hspace{1em} $\vec B_\text{up}$: upsampled sequence
	
	\Statex
	
	\LeftComment{Step 1: Determine which source points are kept}
	\State $\text{idx}_\text{keep}[i] \gets \text{False},\;\forall i=1,\ldots,n_\text{src}-1$
	\State $i\gets1$
	\While{$i< n_\text{src}$}\Comment{point $n_\text{src}$ is always kept, see Step 3}
	\State $\text{idx}_\text{keep}[i]\gets\text{True}$
	\State Draw $g \sim \text{Uniform}\{1, \ldots, \bar{g}\}$
	\State $i\gets i+g$
	\EndWhile
	
	\Statex
	
	\LeftComment{Step 2: Determine how many points are inserted between two source points}
	\State Draw $s \sim \text{Uniform}[0, 0.99]$ \Comment{skewness parameter}
	\State Draw $r[j] \sim \text{Uniform}(0,1),\;\forall j=1,\ldots,n_\text{ins}+1$
	\State $r[j] \gets \epsilon \; \forall\, j \text{ s.t. } r[j] < s$ \Comment{large $s$ forces many $r[j]$ near zero ($\epsilon\gtrapprox0$)}
	\State $r[j] \gets (\sum_{l=1}^{j} r[l]) / (\sum_{l=1}^{n_\text{ins}+1}r[l]), \; \forall\, j=1,\ldots,n_\text{ins}$ \Comment{low $s$: $r$ grows smoothly; high $s$: $r$ grows in steps}
	\State $\text{idx}_\text{ins}[j] \gets \lfloor r[j]\cdot(n_\text{src}-1)\rfloor+1, \; \forall\, j=1,\ldots,n_\text{ins}$
	\State $m_\text{ins}[i] \gets \#\{j : \text{idx}_\text{ins}[j] = i\}, \; \forall\, i=1,\ldots,n_\text{src}-1$ \Comment{number of inserted points in segment $i$}
	\State $m_\text{ins}[i] \gets \bar{m} \; \forall\, i \text{ s.t. } m_\text{ins}[i] > \bar{m}$\Comment{cap insertions per segment}
	
	\Statex
	
	\LeftComment{Step 3: Build upsampled (output) sequence}
	\State $k \gets 1$
	\For{$i=1,\ldots,n_\text{src}-1$}
	\State \algorithmicif~$\text{idx}_\text{keep}[i]$ \algorithmicthen~$\vec B_\text{up}[k]\gets\vec B_\text{src}[i]$, $k \gets k + 1$ \Comment{append source point $i$ if kept}
	\State Draw $\delta[j] \sim \text{Uniform}(0,1),\;\forall j=1,\ldots,m_\text{ins}[i]+1$
	\State $\delta[j] \gets \sum_{l=1}^{j} \delta[l], \; \forall j=1,\ldots,m_\text{ins}[i]+1$
	\State $\delta[j]\gets \delta[j]\cdot\frac{1-(m_\text{ins}[i]+1)\delta_\text{min}}{\delta[m_\text{ins}[i]+1]}+j\cdot\delta_\text{min}, \;\forall j=1,\ldots,m_\text{ins}[i]$ \Comment{enforce minimum spacing}
	\State $\vec B_\text{up}[k+j-1]\gets (1-\delta[j])\cdot\vec B_\text{src}[i]+\delta[j]\cdot\vec B_\text{src}[i+1],\; \forall j=1,\ldots,m_\text{ins}[i]$
	\State $k\gets k+m_\text{ins}[i]$
	\EndFor
	\State $\vec B_\text{up}[k]\gets\vec B_\text{src}[n_\text{src}]$\Comment{last source point is always kept}
\end{algorithmic}
\vspace{0.1cm}
\hrule